\documentclass[12pt]{article}
\usepackage[T1]{fontenc}
\usepackage[utf8]{inputenc}
\usepackage{lmodern}
\usepackage{hyperref}       
\usepackage{url}            
\usepackage{booktabs}       
\usepackage{nicefrac}       
\usepackage{microtype}      
\usepackage{xcolor}         
\usepackage{graphics}       
\usepackage{graphicx}
\usepackage{bm} 
\usepackage{amsmath}
\usepackage{amsfonts}       
\usepackage{mathrsfs}
\usepackage{mathtools}
\usepackage{natbib}
\usepackage{amsthm}
\usepackage{amssymb}
\usepackage{algorithm, algpseudocode}
\usepackage[font=small, labelfont=bf]{caption}
\usepackage[font=small, labelfont=bf]{subcaption}
\usepackage[bottom]{footmisc}
\usepackage{multirow}
\allowdisplaybreaks

\usepackage{enumitem}
\newcommand{\blind}{1}
\date{}

\usepackage{xr}
\makeatletter

\newcommand*{\addFileDependency}[1]{
\typeout{(#1)}
\@addtofilelist{#1}
\IfFileExists{#1}{}{\typeout{No file #1.}}
}\makeatother

\newcommand*{\myexternaldocument}[1]{%
\externaldocument{#1}%
\addFileDependency{#1.tex}%
\addFileDependency{#1.aux}%
}
\myexternaldocument{supplemental-v3}
\usepackage{prettyref}
\newrefformat{S}{~S\ref{#1}}

\usepackage{hyperref}
\hypersetup{colorlinks, citecolor=blue,linkcolor=red}

\mathchardef\mhyphen="2D
\DeclareMathOperator*{\argmin}{argmin}
\DeclareMathOperator*{\argmax}{argmax}

\theoremstyle{exampstyle} \newtheorem{theorem}{Theorem}

\newtheorem{assumption}{Assumption}

\makeatletter
\def\@fnsymbol#1{\ensuremath{\textcolor{black}{\ifcase#1\or *\or \dagger\or \ddagger\or
   \mathsection\or \mathparagraph\or \|\or **\or \dagger\dagger
   \or \ddagger\ddagger\else\@ctrerr\fi}}}
\makeatother

\begin{document}

\def\spacingset#1{\renewcommand{\baselinestretch}%
{#1}\small\normalsize} \spacingset{1}


\if1\blind
{ 
\title{\bf Multi-Task Active Learning with Efficient Resource Allocation}
\author{
Hanwen Ye\thanks{Google, Inc., Mountain View, CA}
 and Jiuchen Zhang\thanks{Department of Biostatistics, University of Michigan, Ann Arbor, MI}
 and  Annie Qu\thanks{Department of Statistics, University of California Santa Barbara, Santa Barbara, CA}
}
{\def\null\vskip 2em{}\maketitle}
\vspace{-1em}
} \fi

\if0\blind
{
  \bigskip
  \bigskip
  \bigskip
  \begin{center}
    {\LARGE\bf Multi-Task Active Learning with Efficient Resource Allocation}
\end{center}
  \medskip
  \bigskip
} \fi

\date{}

\begin{abstract}
    Many scientific studies allow costly auxiliary information to be collected during data labeling but not at deployment. Examples include diagnostic tests, laboratory assays, and expert evaluations. We study prediction under this deployment asymmetry, where auxiliary variables are selectively acquired during labeling under a budget constraint but systematically unavailable at prediction time, creating a missing-by-design problem that couples data acquisition, surrogate construction, and prediction. In this work, we introduce Active Learning with Cost-Adaptive Task Resource Allocations (ALCATRAs), a unified framework for selectively acquiring auxiliary information under resource constraints and leveraging that information to improve downstream prediction. ALCATRAs consists of two main components: a task-selection policy which strategically selects a sequence of cost-effective tasks for unlabeled data to perform, and a surrogate learning procedure which transfers knowledge from completed tasks to enhance model predictions. In theory, we show the effectiveness of surrogate models and sample-efficient task policies in improving the model's prediction error bound. Simulation studies and an application to the UCI heart disease cohort demonstrate improved sample efficiency of the proposed ALCATRAs framework relative to baselines under the studied settings.
\end{abstract}

{\noindent%
}
\vfill

\newpage
\spacingset{1.8} 

\section{Introduction}
\label{sec:intro}

Modern hardware and computing power allow collecting vast amounts of data, which also facilitate the training of large-scale machine learning models for a wide range of prediction tasks, such as image recognition \citep{deng2009imagenet}, stock price forecasting \citep{wang2023knowledge}, and language generation \citep{radford2019language, brown2020language}. However, having large volumes of data alone is not sufficient. The key challenge lies in maintaining high-quality and well-labeled data to achieve accurate machine learning modeling and prediction, which often requires extensive human annotation and with significant costs.   Moreover, labeling is often not a single atomic action: in many practical settings, the analyst can perform additional optional measurements to assist labeling, each carrying its own cost. This creates a need for labeling strategies that not only choose which instances to label but also decide which auxiliary measurements to acquire.

A concrete motivating example, which we revisit in Section~\ref{sec:data} using the UCI heart disease cohort \citep{heart_disease_45}, is the development of a cardiovascular risk-screening model. At enrollment, routine information such as age, sex, chest pain type, resting blood pressure, and maximum heart rate is recorded at no extra cost. Additional diagnostics may also be ordered to refine the clinician's judgment of disease presence: a cholesterol panel, a fasting blood sugar test, a resting electrocardiogram, fluoroscopy of the major vessels, and an exercise stress test, each of which differs in cost by orders of magnitude. A clinical research team building such a screening model from a small labeled cohort therefore operates under a fixed total budget: not every labeled patient can receive every diagnostic, and the investigator must decide, for each patient, which tests to order and in what sequence, stopping once the next test is unlikely to be cost-effective. Once the model is deployed in primary-care or community-clinic settings, the costly diagnostics are no longer part of the screening input, and the predictor must rely on the routinely collected features alone. This structural asymmetry, where auxiliary information is available during labeling but unavailable at prediction time, is the central statistical challenge motivating this work.

Active learning (AL) is an iterative paradigm that selects the most informative unlabeled instance at each step and updates the model after each query \citep{ren2021survey, li2024survey, lughofer2012hybrid, krishnamurthy2019active, deng2023query}. Traditional AL assumes a fixed per-instance labeling cost based on observable features alone. In many practical settings, however, critical information becomes available only after additional resource-allocated tasks are performed. Existing cost-sensitive AL methods \citep{kuo2018cost, tsou2019annotation} have considered varying labeling costs but have not addressed the cost or informational value of these auxiliary tasks.

In this work, we introduce Active Learning with Cost-Adaptive Task Resource Allocations (ALCATRAs), a unified AL framework for settings with supplementary tasks of varying resource requirements. Under a fixed budget, ALCATRAs curates unlabeled data for annotation and sequentially selects cost-effective tasks for each labeled instance. The framework has two components: a task-selection policy and a surrogate learning procedure, both model-agnostic, that jointly assist the predictive model and refine the acquisition strategy across AL iterations.

The contributions of this work are threefold. We first introduce ALCATRAs, a unified framework for active learning under selective acquisition of costly auxiliary variables, where the choice of which auxiliary information to acquire during labeling and the choice of how to transfer that information to a deployment-time predictor are formulated and optimized as a single learning problem under a fixed budget.  Existing feature acquisition methods assume acquired variables remain available at deployment, while existing surrogate learning methods assume auxiliary outcomes are fully observed for every labeled subject. We study the statistically distinct setting where auxiliary variables are selectively acquired during labeling under a budget constraint but are systematically unavailable during deployment. This creates a missing-by-design problem that couples data acquisition, surrogate construction, and prediction, linking the work naturally to two-phase sampling, missing-by-design observation, and adaptive experimental design. Although ALCATRAs contains several algorithmic components, the primary contribution of this paper is the formulation and analysis of active learning under selective auxiliary acquisition and deployment-time unavailability; Streak Learning and Online-ELBO are supporting procedures.

Second, we develop two procedures that operationalize the framework in the small-labeled-data, online-update regime characteristic of active learning. Streak Learning is a sample-efficient acquisition decision rule that targets continuing-optimal decisions early in the acquisition sequence, avoiding the exponential variance growth of standard importance-weighted off-policy estimation \citep{goldberg2012q, zhao2015new, cui2017tree, zhao2020constructing, ye2024stage}. Online-ELBO is a policy-adjusted variational objective that corrects for the selection bias induced by adaptive acquisition and enables surrogate learning when auxiliary outcomes are observed only for a subset of subjects \citep{xia2024prediction, wang2016cost, yin2020reinforcement, lyu2024semi, kallus2024role}.

 Third, we provide supporting theoretical guarantees for the major components of the framework. Under the stated assumptions, a localized excess risk bound shows that surrogate-completed unlabeled data can contract excess-risk convergence by a factor of $(N^U)^{-1/4}$, reducing dependence on labeled samples; finite-sample analyses of the acquisition rule and the variational objective are deferred to the supplementary materials. Simulation studies and an application to the UCI heart disease cohort, complemented by a parallel FashionMNIST analysis in the supplement, demonstrate improved resource efficiency.

The paper is structured as follows. Section~\ref{sec:background} introduces notation and positions ALCATRAs against neighboring frameworks. Section~\ref{sec:method} presents the framework together with the Streak Learning acquisition rule and the online surrogate learning procedure. Section~\ref{sec:theory} provides the theoretical analysis. Section~\ref{sec:simulation} reports simulation results, and Section~\ref{sec:data} applies the framework to the UCI heart disease cohort. Section~\ref{sec:discussion} concludes.

\section{Background} \label{sec:background}

This section introduces the notation used throughout Section~\ref{sec:method} and states the multi-task active learning setting under selective auxiliary acquisition. A short positioning of ALCATRAs against existing feature acquisition and surrogate learning frameworks is given.

\subsection{Notation and Multi-task Setting}

Consider $N$ instances, of which $N^L$ are labeled and $N^U$ are unlabeled. Each instance $i$ has $P$-dimensional observable features $X_i \in \mathcal{X} \subseteq \mathbb{R}^P$ and a labeling target $y_i \in \mathcal{Y}$ (classification or regression). The labeled set is $\mathbb{D}^L=\{X_i, y_i\}_{i \in N^L}$ and the unlabeled set is $\mathbb{D}^U = \{X_i\}_{i \in N^U}$. A prediction model $f \in \mathcal{F}: \mathcal{X} \mapsto \mathcal{Y}$ is trained by empirical risk minimization on $\mathbb{D}^L$,
\begin{equation}
    \hat{f} = \argmin_{f\in \mathcal{F}} \frac{1}{|\mathbb{D}^L|} \sum_{(X,y) \in \mathbb{D}^L}\mathcal{L}(f(X), y),
    \label{eq:AL-goal}
\end{equation}
where the loss $\mathcal{L}$ depends on the target type (e.g., cross-entropy for classification, mean-squared error for regression). Active learning selects new instances to label under a fixed total budget $B \in \mathbb{R}^+$.




Instance selection is guided by a scoring function $U: \mathcal{F} \times \mathcal{X} \mapsto \mathbb{R}$, which quantifies model uncertainty on an unlabeled instance,
\begin{equation}
    X_{\text{new}} = \arg\max_{X \in \mathbb{D}^{U}} U(f, X).
\label{eq:uncertain-select}
\end{equation}

We treat $U$ as given (existing choices include entropy, confidence bounds, and probabilistic scores; see Section~\ref{sec:intro}); our contribution is not in choosing $U$ but in deciding which auxiliary tasks to acquire on the selected instance, as described next.

In the multi-task setting considered in this paper, there exists $Q$ number of supplementary tasks available to assist labeling. Once the $q^{\text{th}}$ task is performed on instance $i$, the task outcome $W_{i,q} \in \mathcal{W}_q \subseteq{\mathbb{R}}$ provides additional covariate information of instance $i$, and incurs a fixed, pre-determined resource cost $c_q \in \mathbb{R}^+$. Note that the task outcome random variable $W_q$ for any given task $q \in \{1,\ldots,Q\}$ could be multi-dimensional; however, for simplicity, we assume it is a real-numbered scalar. Tasks are selected sequentially until all $Q$ tasks have been completed or enough task information is acquired to make a prediction. We denote the decision to perform $q^{th}$ task for instance $i$ at stage $t$ as $a_{i,t}=q$, and the decision to stop acquiring new tasks as $a_{i,t}=0$. Since performing the same task on the same instance yields redundant results, we assume no repeated task performed on any given instance. As a result, the available tasks for instance $i$ at stage $t$ has the form $a_{i,t} \in \mathcal{A}_{i,t} = \{0,1,\ldots,Q\} \setminus\{a_{i,1}, \ldots, a_{i, t-1}\}$, and instances with all tasks completed have a task completion order corresponding to one of the possible permutations, i.e., $(A_{1},\ldots,A_t,\ldots,A_{T=Q}) \in \text{Perm}(\{1,\ldots,Q\})$. Performed tasks are observed in the labeled dataset but no task information is available among the unlabeled dataset. We further denote the labeled data $\mathbb{D}^L$ and unlabeled data $\mathbb{D}^U$ in the multi-task setting as,
\begin{equation}
    \mathbb{D}^L = \{X_i, (W_{i, A_{i,t} \sim \pi_t})_{t=1}^{T_i}, y_i\}_{i \in N^L} \;, \qquad \mathbb{D}^U = \{X_i\}_{i \in N^U},
\end{equation}
where $T_i=\sum_{t=1}^Q \mathbb{I}(a_{i,t} \neq 0) $ represents the number of tasks instance $i$ has completed, and $\pi_t: \mathcal{H}_t \mapsto \mathcal{A}_t$ is the underlying decision policy which determines the observed task completion order from all the historical information $H_t = S_t(X, A_1=a_1, W_{a_1},\ldots,A_{t-1}=a_{t-1}, W_{a_{t-1}})$, summarized by a deterministic function $S_t$ at each time step $t$. As the supplementary tasks require different resources and incur varying costs, in addition to optimizing for the general predictive performance in Equation \eqref{eq:AL-goal}, we aim to identify the optimal task selection policy, $\pi^*$, which maximizes the  potential cost-effectiveness, i.e.,
\begin{equation}
\bm{\pi}^* = \argmax_{\bm{\pi} = (\pi_1,\ldots,\pi_{T=Q})} \mathbb{E}^{\bm{\pi}}\{R\}.
\label{eq:potential-outcome}
\end{equation}
Here, the cost-effectiveness is measured by some underlying rewards $R$. An intuitive formulation of the reward function $R$ would be,
\begin{equation}
    R(W_{\bm{a} \sim \bm{\pi}}; X, y, f,\gamma) = \frac{\text{Information-Gain}(W_{\bm{a}}; X, y, f)}{\gamma \cdot \text{cost}(X, W_{\bm{a}})},
    \label{eq:reward}
\end{equation}
where the information-gain measures the helpfulness of performing selected tasks $W_{\bm{a}}$ for the predictive model $f$ in accurately predicting label $y$ from $X$, and the labor cost includes a fixed base labeling cost $c$ and associated task-specific resource costs, i.e, $\text{cost}(X, W_{\bm{a}}) = c + \sum_{q \in \bm{a}} c_q$. Additionally, $\gamma$ is a hyper-parameter which controls the relative importance of reducing costs compared to the improvement in model performance. The reward in Equation~\eqref{eq:reward} is not the unique choice for measuring cost-effectiveness; alternatives include marginal value of information and expected utility gain. We adopt information-gain per unit cost because it directly operationalizes the resource-allocation objective of maximizing predictive improvement under a fixed budget, and it admits a closed-form empirical estimator (Equation~\eqref{eq:reward-concrete} in Section~\ref{sec:method}) that integrates cleanly with the AL iteration.

\subsection{Related Works: Feature Acquisition and Surrogate Learning}
\label{sec: 2-3}

Two lines of work are most closely related to the multi-task setting above. Feature acquisition methods select additional measurements to improve a predictor, ranging from utility-based imputation \citep{zheng2002active, saar2009active, attenberg2011selective} to reinforcement-learning-based approaches that treat acquisition and prediction jointly within a partially observed Markov decision process \citep{ji2007cost, hausknecht2015deep, igl2018deep, yin2020reinforcement, li2024towards}. These methods assume the acquisition environment remains accessible at prediction time, which does not hold in our deployment-time setting. Surrogate learning \citep{sinha2019variational, kallus2020role, xia2024prediction} addresses the missing auxiliary outcomes at prediction time by training a surrogate predictor on the labeled set, but existing methods assume the auxiliary outcomes are fully observed for every labeled subject, whereas in our setting they are selectively acquired through an evolving policy. ALCATRAs is designed to handle both the missing-at-deployment and missing-by-design structures jointly. The setting also connects to several adjacent statistical literatures: two-phase sampling designs \citep{breslow1999design} in which expensive auxiliary measurements are obtained only on a phase-II subsample, missing-by-design observation \citep{robins1994estimation} in which the missingness mechanism is governed by a known, controllable acquisition rule, and adaptive experimental design \citep{pukelsheim2006optimal, thompson1990adaptive} in which measurement decisions are part of the inferential procedure. Standard active learning, by contrast, assumes that information acquired during labeling is identical to information available during deployment. Our setting introduces a systematic discrepancy between the two phases, creating a missing-by-design problem that cannot be addressed by classical active learning alone. Table~\ref{tab:positioning} summarizes these structural distinctions.

\begin{table}[h]
\centering
\caption{Positioning of ALCATRAs against neighboring frameworks.}
\label{tab:positioning}
\resizebox{\textwidth}{!}{%
\begin{tabular}{lccc}
\toprule
Framework & Auxiliary variables & Sequential acquisition & Available at deployment \\
\midrule
Active Learning            & No   & No   & N/A                  \\
Feature Acquisition (POMDP) & Yes  & Yes  & Required             \\
Surrogate Learning         & Yes  & No (one-shot) & Not required \\
\textbf{ALCATRAs (this work)} & Yes & Yes & Not required \\
\bottomrule
\end{tabular}}
\end{table}

\section{Methodology} \label{sec:method}

In this section, we formally present ALCATRAs framework to optimize task allocation and improve model predictive performance in the multi-task AL setting. The generic framework is model-agnostic, allowing flexbility in the choice of task-selection policies and surrogate learning procedures. To further improve adaptability of the proposed framework, we introduce a task-acquisition procedure, Streak Learning, designed for sample efficiency by encouraging continuing-optimal decision sequences, and an online surrogate learning procedure which dynamically refines task outcome representations as the task policy evolves.

\subsection{ALCATRAs: Cost-Adaptive Task Resource Allocations}

\begin{figure}[h]
    \centering
    \includegraphics[width=0.7\linewidth]{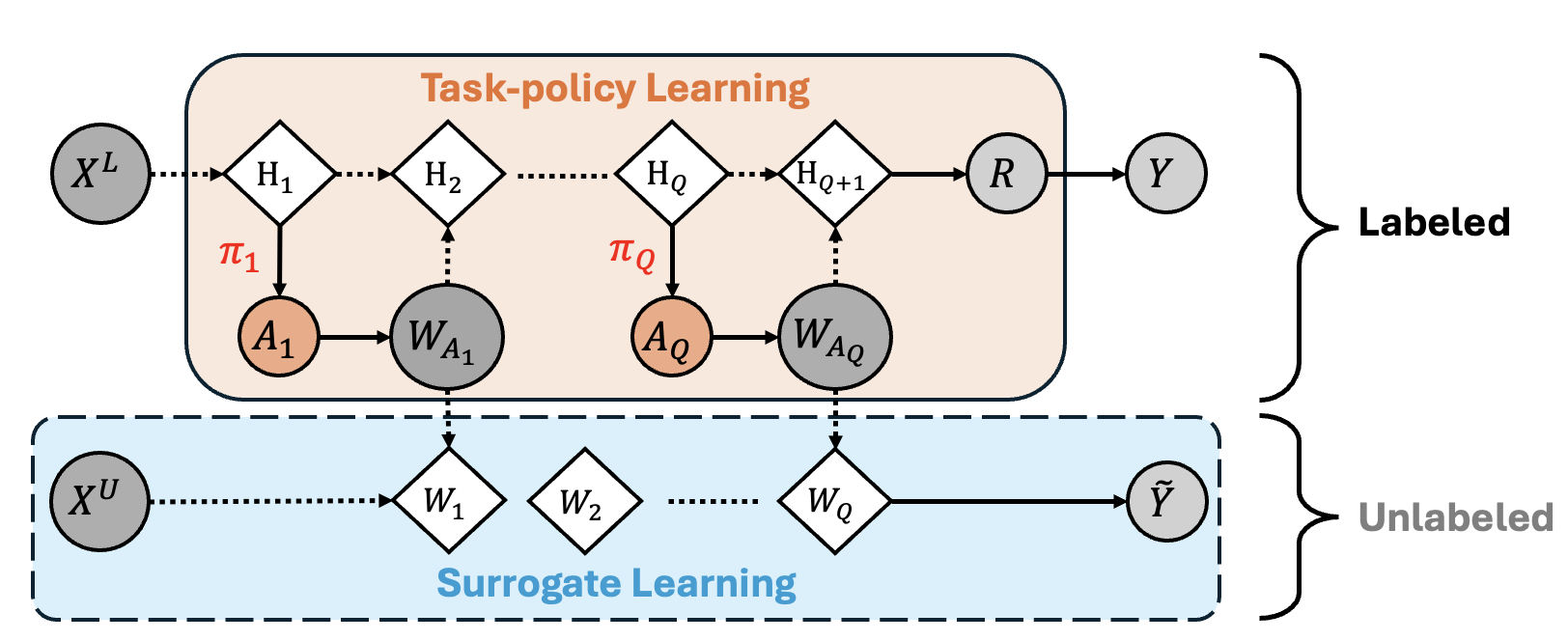}
    \caption{Graphical model for task-policy and surrogate learning processes within one AL iteration. Task-policy is learned among the labeled dataset, whereas surrogate learning estimates surrogate labels for the unlabeled dataset. Shaded nodes denote observable variables, while white diamond-shaped nodes represent hidden states or variables requiring imputation. Dashed arrows indicate inference steps, and solid arrows represent direct computations.}
    \label{fig:graphical-model}
\end{figure}

Supplementary tasks assist labeling but are limited to the labeling stage, as budget constraints make it impractical to perform tasks for all unlabeled instances. ALCATRAs accounts for this dynamic availability of task resources through a two-step learning procedure. First, during the labeling stage, ALCATRAs selects the most informative unlabeled instance $X_{\text{new}}$ to label using a scoring function $U$ in Equation \eqref{eq:uncertain-select}, and determines a sequence of supplementary tasks $\{A_{\text{new}, t}\}_{t=1}^{T_{\text{new}}}$ to perform based on a task policy $\bm{\pi}=(\pi_1,\ldots,\pi_Q)$ i.e.,
\begin{equation}
    A_{\text{new}, t} = \pi_t (H_{\text{new}, t})\;\text{for } 1 \le t \le T_{\text{new}} \le Q.
    \label{eq:task-selection}
\end{equation}
The number of tasks $T_{\text{new}}$ to complete varies across each instance and is determined by the policy. Once the requested task information is acquired, in the second step, ALCATRAs updates the prediction function $f$ and task-selection policy $\pi$ using pre-defined surrogate learning procedure and task-policy learning procedure with newly labeled data
$\mathbb{D}^L \;\cup\; \{X_{\text{new}}, \{W_{\text{new}, A_{\text{new}, t}}\}_{t=1}^{T_{\text{new}}}, y_{\text{new}}\}$. An illustration of ALCATRAs is shown in Figure \ref{fig:graphical-model}.

Compared to classic AL, ALCATRAs does not simply introduce a task-selection step during the labeling stage in Equation \eqref{eq:task-selection}. Instead, the proposed framework integrates supplementary tasks throughout the entire AL iterative process, including how to select more cost-effective tasks and leverage their outcomes for future predictions. These two objectives are mutually dependent and achieved by two learning procedures, task-policy learning and surrogate learning procedure respectively. Both procedures are flexible in terms of model specifications. The task-selection policy can be estimated using any finite-horizon decision-making algorithm to maximize the reward defined in Equation \eqref{eq:potential-outcome}. Meanwhile, the surrogate learning can employ an arbitrary estimation procedure to infer surrogate labels $\tilde{y}^U$ for unlabeled instances by leveraging all previously observed task information, i.e.,
\begin{equation}
\tilde{y}^U = g(X^U, \hat{W}^U; X^L, W^L).
\label{eq: surr-g}
\end{equation}
where $g \in \mathcal{G}: \mathcal{X \times\mathcal{W \mapsto \mathcal{Y}}}$ is a surrogate predictive function, incorporating task outcomes as an additional predictor. The surrogate labels $\tilde{y}^U$, containing task information, are then used to update the target predictive model,
\begin{equation}
\hat{f} = \argmin_{f \in \mathcal{F}} \mathbb{E}_{|\mathbb{D}^L|} \{\mathcal{L}(f(X^L), y^L)\} + \mathbb{E}_{|\mathbb{D}^U|} \{\mathcal{L}(f(X^U), \tilde{y}^U)\}.
\label{eq: surr-eq}
\end{equation}
As noticed, the predictive model $f$ benefits from the task information through the surrogate labels, without directly relying on tasks outcomes as predictors, which are completely missing among unlabeled instances.

Once the predictive model is updated from $f^{(t)}$ to $f^{(t+1)}$ after one AL iteration, we evaluate the cost-effectiveness of selected tasks by measuring the model information gain achieved by the newly labeled data and observed task outcomes. The corresponding reward for past task selections and policy learning in Equation \eqref{eq:reward} can be explicitly computed as the reduction in uncertainty scoring, i.e., 
\begin{equation}
    R_{\text{new}} = \frac{U(f^{(t)}, X_{\text{new}}) - U(f^{(t+1)}, X_{\text{new}})}{\gamma \cdot \text{cost}(X_{\text{new}}, W_{\text{new}, \bm{a}_{\text{new}}})}.
\label{eq:reward-concrete}
\end{equation}
With the updated reward, the task-selection policy can be optimized to choose more effective tasks in subsequent labeling stages. In turn, these improved task selections contain more valuable task information for surrogate learning to extract and refine its surrogate label predictions as iterations progress. Together, these two learning processes iteratively improve predictive performance and enable a more efficient allocation of labeling task resources within the integrated ALCATRAs framework. 

The proposed framework offers a high degree of flexibility, allowing the integration of various task policies and surrogate learning algorithms. However, the choice of these algorithms can significantly affect the framework’s overall performance. For instance, a randomly selected task policy is unlikely to achieve the same level of cost-effectiveness as a carefully optimized task policy. Similarly, the surrogate learning may fail to generate accurate surrogate labels if the task representations for unlabeled instances are not well-constructed. In the following, we accompany ALCATRAs with a sample-efficient task-policy learning strategy and an online surrogate learning procedure to enhance the proposed framework's adaptability. These components are specifically designed to accommodate the initially small labeled dataset and the iterative nature of AL.

\subsection{Streak Learning for Task Selections}

Off-policy estimation of the value function in Equation~\eqref{eq:potential-outcome} requires reweighting the rewards observed under a behavioral policy $\bm{\pi}^0$ to the target policy $\bm{\pi}$. The natural estimator is full inverse-propensity weighting (IPW), in which every trajectory is reweighted by the product of stagewise importance ratios $\prod_{t=1}^{T} \pi_t(A_t \mid H_t) / \pi_t^0(A_t \mid H_t)$. In the small-labeled-data regime of active learning, full IPW is statistically infeasible: the product of $T$ stagewise ratios has variance that grows multiplicatively with the decision horizon, a phenomenon known as the curse of horizon. The motivation for Streak Learning is to escape this variance growth by performing partial trajectory matching: only the prefix of decisions that agree with the target policy is reweighted, and stages after the first disagreement are downweighted rather than multiplied through. This relaxation trades a small bias for a substantial reduction in estimator variance, and is what enables policy search on labeled cohorts of practical size.

 Under the standard Positivity, Sequential Ignorability, and Stable Unit Treatment Value (SUTVA) assumptions detailed in the Supplementary Materials, this partial-matching idea is formalized by the Streak Learning estimator for the value function in Equation~\eqref{eq:potential-outcome},
\begin{align}
    \bm{\pi}_{\text{Streak}}^* &= \argmax_{(\pi_1, \ldots,\pi_{Q}) \in \bm{\pi}} \text{VF}^{\text{Streak}}(\pi_1, \ldots,\pi_{Q}) \notag \\
    &= \argmax \mathbb{E} \left\{ \sum_{\delta=1}^{\bar{T}+1} \phi(\delta) \cdot \frac{R \cdot {\prod_{t=1}^{\delta-1}} \;\mathbb{I}(\pi_t(H_t)=A_t) \cdot {\sum_{t=\delta}^{\bar{T}}} \;\mathbb{I}(\pi_t(H_t)=A_t)))}{\prod_{t=1}^T \pi^0_t(A_t | H_t)}\right\}.
    \label{eq: SL-VF}
\end{align}
Specifically, $\bm{\pi}^0$ represents the behavioral policy, which determines the selection probabilities of observed tasks. $\bar{T} \in \{0,1,\ldots,Q\}$ is a random variable indicating the number of completed tasks in a labeling process. $\delta$ denotes the first decision stage where the selected task is not underlying optimal. In other words, $\delta-1$ represents the number of consecutive (streak) tasks selected optimally from the beginning of the sequence. Additionally, $\phi(\cdot)$ is a scaling function specifying the weight placed on maintaining a streak of optimal decisions. Streak Learning may be viewed as a structured partial-matching estimator that interpolates between full trajectory matching (which incurs the variance explosion of full inverse propensity weighting) and stagewise matching (which is biased), thereby trading a small amount of bias for substantial variance reduction.

The proposed Streak Learning method has the following two advantages. First, our estimator naturally adapts to observed task sequences of varying lengths by incorporating the task sequence length as a random variable, $\bar{T}$. In practical labeling processes, certain basic variables may already contain sufficient label information, making it unnecessary to complete all supplementary labeling tasks. This leads to imbalanced task sequences, which is challenging for typical direct-search estimators assuming a fixed decision horizon. In contrast, the Streak Learning estimator evaluates each decision-length scenario and determines, at each stage, whether it is more beneficial to stop early instead of continuing requesting tasks and allocating resources. This adaptive mechanism of early-stopping avoids performing unnecessary tasks which would only provide minimal benefits to the predictive model, thereby minimizing task resource usage.

\begin{figure}[H]
    \centering
    \includegraphics[width=0.78\linewidth]{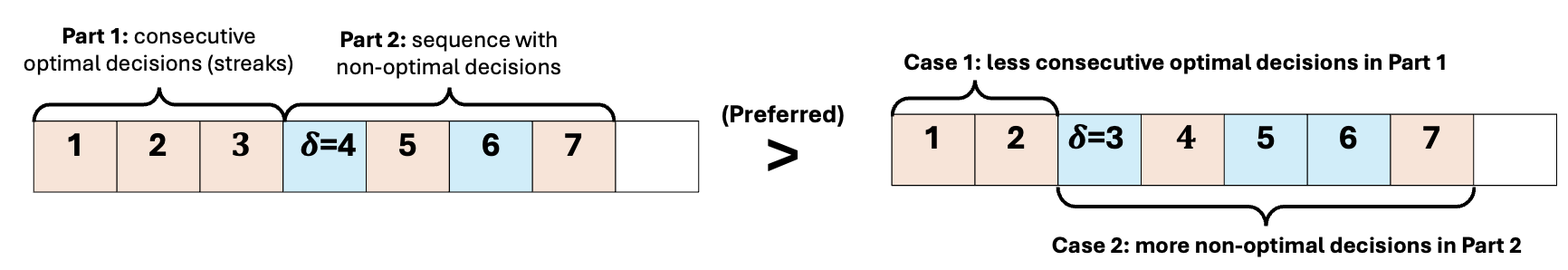}
    \caption{An illustrative example of a two-part decision sequence, where $\bar{T}=7$ out of $Q=8$ available tasks are completed. The pink regions represent the tasks are assigned optimally, while the blue regions indicate tasks are non-optimal.}
    \label{fig:streak}
\end{figure}

Secondly, the proposed estimator improves both sample and cost efficiency by imposing a unique structure on the task selection sequence. Specifically, Streak Learning divides the entire decision sequence into two parts, separated by the decision stage $\delta$. The first part consists of consecutive optimal decisions (streaks), while the second part includes task selections with non-optimal decisions. Figure \ref{fig:streak} illustrates this separation, i.e.,
\noindent The key observation is that, once a non-optimal task is selected in a decision sequence, subsequent selections from the remaining tasks are more likely to result in further non-optimal decisions, increasing task costs. Consequently, improving cost efficiency can be viewed as maintaining a streak of optimal decisions from the beginning of the decision process. Meanwhile, if a non-optimal decision is unfortunately made, the objective shifts to reduce the possibilities of making future non-optimal decisions as much as possible. Ultimately, we aim to align the recommended task allocations to the underlying optimal sequence order. In addition, by accounting for every possible decision stages at which a non-optimal decision could occur, i.e., $\delta \in \{1,\ldots,\bar{T}+1\}$, Streak Learning searches for the optimal decision sequence across all possible combinations of decisions. Compared with common finite-horizon policy-search methods \citep{goldberg2012q, zhao2015new, zhao2020constructing}, which rely on inverse-probability weighting and therefore suffer from the \textit{curse of full-matching} discussed in \cite{ye2024stage}, our method fully leverages the observed task data and substantially improves sample efficiency.

To this point, we provide a tailored sample-efficient task-policy learning procedure within the ALCATRAs framework. In the following subsection, we introduce an online surrogate learning procedure to enhance task outcome representations and facilitate the generation of surrogate labels which align with the evolving task policy.

\subsection{Surrogate Learning with Online Task Representations}

To effectively infer surrogate labels for unlabeled instances using task-based information, we present a policy-adaptive variational inference framework. Specifically, we assume the existence of a latent variable $Z_q \in \mathcal{Z}_q$, which can fully represent the task outcomes such that $W_{q}$ is generated via a pre-specified conditional distribution $F_q(W_q | Z_q)$ with location $\mu$ and scale $\sigma$ parametrized by $Z_q$, i.e., $W_q \;|\; Z_q \sim F_q(\mu(Z_q), \sigma(Z_q))$.

The primary objective is to learn the posterior $P(Z_q | W_q)$ from the labeled dataset and subsequently use it to make inference on future task outcomes for the unlabeled set through the generative model $P(W_q | Z_q)$ derived from $F_q$. Since the true posterior is intractable, we approximate it using a variational distribution $Q_\phi(Z_q | X)$, parameterized by $\phi$. Crucially, to learn $\phi$ consistently within an interactive environment where the data collection mechanism evolves with the task-selection policy, we introduce  a policy-adjusted Evidence Lower Bound (Online-ELBO) that integrates inverse probability weighting (IPW) into the standard ELBO to correct for the selection bias induced by the acquisition policy. Within the selective acquisition framework, the policy-adjusted ELBO serves as the bias-correction component that allows surrogate learning to proceed on adaptively acquired data, i.e.,
\begin{align}
    \text{ELBO}_{\text{online}} &= \mathbb{E}_{Q_{\phi}(Z_q|X)} \{\mathbb{E}_{W_{q \sim \pi}}\{\log P(W_q|Z_q) \;|\; \pi\}\} - \mathbb{KL}\{Q_{\phi}(Z_q|X) \;||\;p(Z_q)\} \\
    &=\mathbb{E}_{Q_{\phi}(Z_q|X)} \left\{\frac{1}{\pi(A=q  \;|\; X)} \cdot \log P(W_q|Z_q) \right\} - \mathbb{KL}\{Q_{\phi}(Z_q|X) \;||\;p(Z_q)\}.
\label{eq: online-ELBO} 
\end{align}
Once the variational parameters $\phi$ are optimized via maximizing our bias-corrected objective, we address the structural missingness in the unlabeled set. Specifically, we utilize the learned distribution to sample latent representations $\hat{Z}^U \sim Q_\phi(\cdot | X^U)$ and subsequently impute the unobserved task features $\hat{W}^U$ using the generative model $F_q$. These imputed features are then incorporated into the surrogate model from Equation \eqref{eq: surr-g} to yield information-rich surrogate labels. Importantly, this procedure enables the target predictive model to incorporate auxiliary task information without incurring the cost of physically performing tasks on the unlabeled population.

We summarize two key advantages provided by our online-ELBO. First, through inversely weighting the task generative model with the task-selection probability, the objective bound prioritizes tasks which are less frequently selected. Due to the evolving task-selection policy, the observed task outcomes do not form a representative sample of the overall task distribution. Without proper adjustment, the learned task representation from standard variational learning \citep{kingma2013auto} may overfit to frequently selected tasks while underrepresenting ones which are more difficult to observed. By leveraging selection probabilities, we can identify and assign greater weights to underrepresented task outcomes, improving their reconstruction accuracy and leading to a more robust task representation.

The weighting mechanism also seamlessly connects the surrogate learning process to the task-selection procedure within the integrated ALCATRAs framework. At each AL iteration, the task-selection policy and corresponding selection probabilities are explicitly known, which allows an unbiased estimation of the underlying task outcome distribution without additional selection probability estimation. Moreover, task representations learned from this unbiased distribution are more robust and can potentially improve the accuracy of surrogate label predictions, resulting into a more precise assessment of model performance gains in Equation \eqref{eq:reward-concrete} by mitigating surrogate prediction errors. Importantly, these performance gains, which reflects the impact of selected tasks on improving the predictive model, subsequently guide and refine the task-selection policy in future AL iterations.

Under positivity and sequential ignorability (Assumptions~\ref{A-ignorability} to~\ref{A-pos} in the Supplementary Materials), the proposed Online-ELBO is causally unbiased for the population Evidence Lower Bound of the task outcome distribution, $\mathbb{E}_{X, A, W_A} [\mathrm{ELBO}_{\mathrm{online}}] = \mathrm{ELBO}_{\mathrm{population}}$. The formal statement and proof are given as Theorem~\ref{thm:elbo-unbiased} in the Supplementary Materials.

In conclusion, we introduce an online variational surrogate learning method tailored to the ALCATRAs framework. Our approach learns a latent task representation to fully describe the distribution of completely missing task features in unlabeled instances. This makes it possible to predict surrogate labels for the unlabeled with more informative task outcomes than observable features alone. By leveraging inverse weighting, we dynamically adapt the imputation process to the evolving task-selection policy, which improves sample efficiency for estimating an unbiased task outcome distribution. The complete iterative procedure, parametrization choices, and optimization details are given as Algorithm~\ref{alg: AL-1} in Section~\ref{chapter3-appendix: implementation} of the Supplementary Materials. The next section presents the theoretical results, along with detailed analysis of task-selection and surrogate learning procedures.

\section{Theoretical Results} \label{sec:theory}

In this section, we establish theoretical results for the target predictive model, sequential task-acquisition policy, and variational outcome representations to show the convergence and cost efficiency guarantees of our multi-task ALCATRAs framework. Detailed assumptions, proofs, and auxiliary lemmas are deferred to the Supplementary Materials.

\subsection{Surrogate Generalization and Localized Excess Risk Bound}

 We now state the main risk-control result for ALCATRAs. The excess risk of the target predictor is $R(\hat{f}) - R(f^*)$, where $R(f) = \mathbb{E}\{(f(X) - y)^2\}$ is the population risk. Because standard risk bounds over the entire class $\mathcal{F}$ are typically loose, we restrict our complexity analysis to a localized function class $\mathcal{F}_\epsilon \coloneqq \{f \in \mathcal{F}: \|f - \tilde{g}\|_U \le \epsilon\}$ around the empirical online surrogate predictor $\tilde{g}(X) = \hat{g}(X, \hat{W}(X))$, which maps observable features and imputed task outcomes to target predictions. 

\begin{assumption} (Localized Target Realizability)\label{ass:realizability_main}
Both the population optimal target predictor $f^*$ and constrained empirical surrogate estimator $\hat{f}$ reside in the restricted localized class $\mathcal{F}_\epsilon$.
\end{assumption}

  This assumption formalizes settings in which the surrogate predictor captures most of the predictive information contained in the selectively acquired auxiliary variables: $\tilde{g}$ uses the same target labels but is trained on the richer information set $(X, W)$ rather than $X$ alone, so candidate models for the deployment-time predictor $\hat{f}$ are expected to lie close to $\tilde{g}$ on the unlabeled pool. We now state the main result.

\begin{theorem} (Localized Excess Risk Bound for Surrogate Learning)\label{thm:main}
By Assumptions \ref{ass:realizability_main} and regularity assumptions \ref{ass:boundedness}--\ref{ass:realizability} in Supplemental Materials, for every $x > 0$, with probability at least $1 - e^{-x}$, the localized excess risk of the target predictive model is bounded by:
\begin{align}
    R(\hat{f}) - R(f^*) &\le \underbrace{2\epsilon^2 + C_1 \left( \frac{\epsilon^{1-\alpha_{\mathcal{F}}/2}}{\sqrt{N^L}} \right)}_{\text{Localized Complexity}} + 8B \sqrt{x + \ln(6)} \Bigg( \underbrace{\frac{\epsilon + \sqrt{2} L_g \|W - \hat{W}\|_L + \xi}{\sqrt{N^L}} }_{\text{Surrogate Complexity}} \notag \\
    &\quad + \underbrace{\sqrt{C_2} \frac{\epsilon^{1/2 - \alpha_{\mathcal{F}}/4}}{(N^U)^{1/4} \sqrt{N^L}}}_{\text{Unlabeled Contraction}} + \underbrace{\frac{\sqrt{C_3} Q^{1/4} + \sqrt{C_4}}{(N^L)^{3/4}}}_{\text{Labeled Sample Complexity}}\Bigg) + C_5 \frac{x + \ln(6)}{N^L},
\end{align}
where $\xi = \|g^* - y\|_2$ is the irreducible surrogate error, $\alpha_\mathcal{F}\in (0,2)$ quantifies the metric entropy of function class $\mathcal{F}$, and $C_1, C_2, C_3, C_4, C_5 > 0$ are finite constants depending only on the Lipschitz continuity and boundedness parameters of the respective function classes.
\end{theorem}

Theorem~\ref{thm:main} decomposes the excess target risk into localized complexities, surrogate learning errors, and unlabeled contraction rates. The localized complexity term represents the bias-variance tradeoff of restricting our search to a smaller function space. By reducing the function complexity ($\alpha_{\mathcal{F}} \rightarrow 0$) or shrinking the localization radius ($\epsilon \rightarrow 0$) as the labeled sample size grows, we minimize the approximation bias and accelerate excess risk convergence. Meanwhile, the surrogate complexity incorporates the variational representation error into the final target predictive risk. As this representation error decays under increasing labeled samples, the surrogate prediction error remains bounded, ensuring that the target excess risk converges even when task outcomes are structurally missing for unlabeled instances. Furthermore, the essence of surrogate learning is captured by the unlabeled contraction term, which leverages massive unlabeled observations to shrink the target complexity space by a factor of $(N^U)^{-1/4}$ and directly accelerates target domain convergence.

From a practical standpoint, Theorem~\ref{thm:main} suggests that surrogate-assisted active learning benefits from two mechanisms: (i) reduced effective complexity through localization around the surrogate predictor, and (ii) contraction through the unlabeled sample size $N^U$, which is unavailable to standard active learning procedures relying solely on labeled observations.

In summary, while traditional active learning restricted to the labeled dataset converges at the standard rate of $\mathcal{O}(1/\sqrt{N^L})$, our framework utilizes massive unlabeled data ($N^U \gg N^L$) and further contracts the excess risk bound by the factor of $(N^U)^{-1/4}$.   One implication of Theorem~\ref{thm:main} is that ALCATRAs can leverage the unlabeled $X$-only sample to contract the excess-risk bound by an additional $(N^U)^{-1/4}$ factor under the stated assumptions. The result also depends on a stable surrogate completion $\tilde{g}(X)$, which the Online-ELBO objective of Section~\ref{sec:method} provides as an unbiased estimator of the population variational objective (Theorem~\ref{thm:elbo-unbiased} in the Supplementary Materials). In the cardiovascular screening example of Section~\ref{sec:intro}, the practical implication is that a small labeled cohort with selectively run diagnostics can support a predictor that generalizes across the much larger pool of patients seen with only routine measurements, which is exactly the deployment-time setting that distinguishes our problem from standard active learning and feature acquisition.


\subsection{Sample and Cost Efficiency of Streak Learning}

Off-policy sequential decision optimization under active decision processes is highly susceptible to selection bias and extreme variance, making policy search practically impossible under constrained active learning budgets. To justify our sequential task allocation optimization, we show the Fisher consistency of Streak Learning,

\begin{theorem}[Fisher Consistency of Streak Learning]\label{thm:fisher}
Under causal regularity Assumptions \ref{A-SUTVA}--\ref{A-pos} in Supplementary Materials, if the weighting function satisfies $\phi(\delta) > 0$ for all $\delta \in \{1, \ldots, Q+1\}$, the optimal policy $\bm{\pi}^*$ is the unique maximizer of the Streak Value Function:
\begin{equation}
    \bm{\pi}^* = \argmax_{\bm{\pi}} \text{VF}^{\text{Streak}}(\bm{\pi}).
\end{equation}
\end{theorem}

This guarantees that optimizing the Streak Value Function remains asymptotically unbiased.  To characterize the finite-sample efficiency of Streak Learning relative to a trajectory-matching benchmark, we analyze the off-policy variance dominance of our estimator:

\begin{theorem}[Variance Dominance of Streak Learning]\label{thm:variance_reduction_main}
Under causal regularity Assumptions \ref{A-SUTVA}--\ref{A-pos} in Supplementary Materials, for any weighting function $\phi(t) > 0$, the variance of the Streak Learning estimator is strictly dominated by the standard IPW estimator:
\begin{equation}
    \underbrace{\text{Var}\left( \widehat{\text{VF}}^{\text{Streak}}(\bm{\pi}) \right) \le \mathcal{O}\left( \frac{B_R^2 Q^2 \phi(\bar{\delta})}{N^L \epsilon_0^{\bar{\delta}}} \right)}_{\text{Streak Learning (Horizon-Escaping)}} \ll \underbrace{\text{Var}\left( \widehat{\text{VF}}^{\text{IPW}}(\bm{\pi}) \right) \asymp \mathcal{O}\left( \frac{B_R^2}{N^L \epsilon_0^Q} \right)}_{\text{Standard IPW (Curse of Horizon)}},
\end{equation}
where $\bar{\delta} \coloneqq \mathbb{E}[\delta_i] \in [1, Q+1]$ is the expected policy deviation depth.
\end{theorem}

 Theorem~\ref{thm:variance_reduction_main} establishes a finite-sample efficiency advantage of Streak Learning over conventional IPW under the stated assumptions. In standard IPW, matching the entire target decision sequence requires dividing by behavioral policy joint probabilities across all stages, causing an exponential variance explosion $\mathcal{O}(\epsilon_0^{-Q})$ (the curse of horizon). In contrast, Streak OPE restricts sequential importance weight divisions strictly to stages $t < \delta$ (before the optimal streak breaks), completely avoiding probability divisions after the first non-optimal decision. This contracts the exponential penalty strictly to the expected deviation depth $\bar{\delta} \ll Q$.  Under early policy divergence, this variance reduction can be exponential, supporting policy learning with fewer labels in settings where the stated positivity and ignorability conditions hold.

\section{Simulation} \label{sec:simulation}

In this section, we present extensive simulation results to demonstrate the superior sample efficiency of the proposed ALCATRAs compared to several popular active learning frameworks, including classic uncertain-sampling AL \citep{lewis1995sequential}, cost-sensitive AL \citep{kuo2018cost}, AL with generative labels (surrogate-learning) \citep{wang2016cost}, and active feature acquisition with reinforcement learning \citep{yin2020reinforcement}. Due to framework limitations described in Section \ref{sec: 2-3}, only feature acquisition and ALCATRAs are able to leverage the additional task features. The main objective of this simulation is to investigate how the predictive model improves under a fixed budget with access to additional task features, and to evaluate the cost efficiency of selecting these supplementary task features.

The detailed simulation setting is described as follows. First, we independently sample $P$ number of basic covariates $X$ from a standard multivariate normal distribution $\mathcal{N}(0, I_p)$, and then construct each of the $Q$ task features, $W_q$, using a nonlinear transformation of $X$ along with some unobservable noise, i.e, $W_q \sim \mathcal{N}( f^W_{q} (X_{1:P}), \epsilon_q)$, where $\epsilon_q \sim \text{Unif}(1, 5)$ represents the variance of the unobservable noise and $f_q$ is a complex nonlinear function randomly generated from a basis of functions $\{X, X^2, X^3, \arctan X, \text{sign}(X)\}$. Intuitively, the magnitude of the task variance $\epsilon_q$ indicates the inherent uncertainty of the task and could serve as a criterion for task selection. However, to introduce additional complexity to the task acquisition process, we define an optimal task-selection policy which specifies the oracle task acquisition priority as follows,
\begin{equation}
    \pi^*_{t+1}(X, W_{A_{1:t}}) = \argmax_{q: \; 0 \le q \le Q} \; \text{Availability}(q; A_{1:t}) \cdot \text{Score}(q; X, W_{A_{1:t}}).
\end{equation}
Specifically, the availability term, defined as  $\mathbb{I}(q \notin A_{1:t}) \cdot \mathbb{I}(0 \notin A_{1:t})$, ensures that the task $q$ has not been previously selected and the acquisition stopping signal $q=0$ has not been triggered. Meanwhile, the score, obtained from a nonlinear function $f^\pi_{t+1,q}(X, W_{A_{1:t}})$, reflects the relative importance and selection priority of $q^{\text{th}}$ task at stage $t+1$. As a result, the optimal acquisition order is $A^*_{1:\Tilde{T}^*} = \{\pi^*_1(X), \pi^*_2(X, W_{A^*_{1}}), \ldots, \pi^*_{\Tilde{T}^*}(X, W_{A^*_{1:\Tilde{T}^*-1}})\}$, where $\Tilde{T}^*$ is the optimal number of tasks to be completed (i.e., $A_{\Tilde{T}^*} = 0$ if $\Tilde{T}^* < Q$).

Next, given the optimal task acquisition order, we aim to bridge the target regression outcome $y$ to the oracle selection order. Ideally, task features which are first selected should be the most informative for predicting 
$y$ and thus have the largest feature importance. Inspired by the idea of principal components, we define the regression target $y$ as a linear combination of acquired task features with their loadings determined by the selection order,
\begin{equation}
    y = f^y(X) + \sum_{t=1}^{\Tilde{T}^*} (1-\frac{t}{Q+1}) \cdot W_{A^*_{t}} + \epsilon_y,
\end{equation}
where $f^y: \mathbb{R}^P \mapsto \mathbb{R}$ is a nonlinear function which accounts for the main effect of basic features $X$, and $\epsilon_y \sim \mathcal{N}(0,1)$ is independent standard normal noise.

Lastly, we specify the labor costs under the multi-task AL framework. The total cost decomposes into three additive components: a base labeling cost $f^c(X)$ for instance $X$, a task acquisition cost $\sum_{q \in A_{1:\Tilde{T}}} c_q$ over the selected tasks with per-task cost $c_q \sim \text{Unif}(0,1)$, and an efficiency cost that compares the acquired sequence $A_{1:\Tilde{T}}$ to the optimal sequence $A^*_{1:\Tilde{T}^*}$. The efficiency cost itself has two terms: a ranking penalty $\alpha$ that grows when high-priority tasks are acquired after lower-priority ones, and an order penalty $\beta$ that grows with strict deviation of the prefix of $A_{1:\Tilde{T}}$ from $A^*_{1:\Tilde{T}^*}$, with $\alpha, \beta \in [0,1]$ controlling the relative weight. The closed-form expressions for both penalty terms, including the indicator-based ranking comparator and prefix-match formula, are given in Supplementary Section~\ref{chapter3-appendix: sim-cost}. The full cost model therefore acts as both the budget consumed by acquisition decisions and a quality metric for the task acquisition order.

Under the specified simulation setting,  we first train a predictive model $f: \mathcal{X} \mapsto \mathcal{Y}$ on a labeled dataset $\mathbb{D}^L = \{X, W_{A \sim \pi_0}, y\}$ of size $N^L$, where the task features are observed according to a random behavior policy $\pi_0$. Then, we allow the AL framework to actively query labels and acquire task features from an oracle unlabeled dataset $\mathbb{D}^U = \{X, W_{1:Q}, y\}$ of size $N^U$ until a fixed budget $B$ is exhausted. At each AL iteration, the empirical mean-squared error (MSE) on a held-out testing dataset, $\frac{1}{|\mathbb{D}^{\text{te}}|}\sum_{(X,y) \in \mathbb{D}^{\text{te}}}\{(f(X) - y)^2\}$, is recorded to evaluate the evolving predictive performance of $f$. We repeat the AL process 100 times for each parameter setting and report the evaluation metrics below. 

\subsection{Budget‐Constrained Convergence of AL Frameworks}

In this simulation, we aim to visualize how the predictive function evolves as more unlabeled data are beling labeled. In addition, since labeling instances and completing supplementary tasks incur varying costs, to emphasis the role of labeling costs, we present the change in model predictive performance with respect to cost consumption percentages.

\begin{figure}
    \centering
    \includegraphics[width=0.8\linewidth]{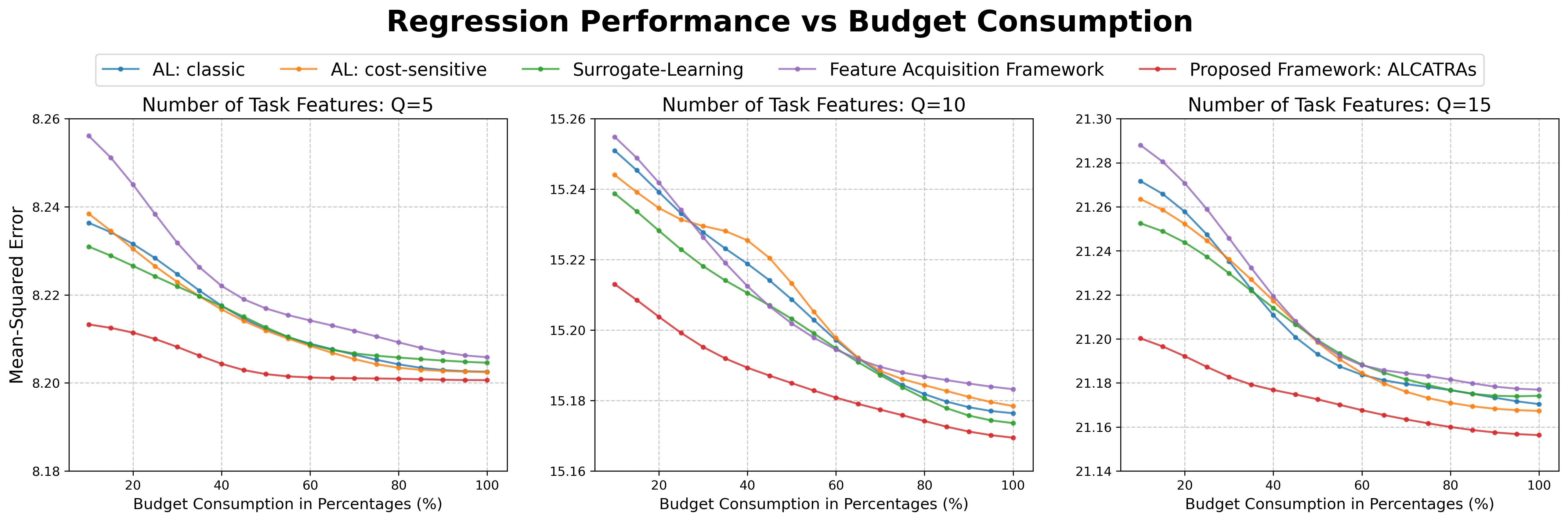}
    \caption{Trajectories of model predictive performance in MSEs versus the budget consumption when the initial labeled data size $N^L=500$, unlabeled data size $N^U=1000$, and maximum budget $B=500$. A 100\%  budget consumption means the entire budget of 500 has been fully utilized.}
    \label{fig:sim-framework}
\end{figure}

As shown in Figure \ref{fig:sim-framework}, across all scenarios with varying numbers of tasks, the proposed ALCATRAs achieves the fastest convergence with the least amount of budget consumption, and  improves over baselines in both initial and overall predictive performance across the studied settings. Moreover, as the number of task features increases, the advantage of ALCATRAs becomes even more pronounced. There are two main factors contributing to this performance gain. First, similar to surrogate learning, the use of surrogate labels provides an initial performance gain by effectively increasing the amount of training data. By leveraging additional task features, ALCATRAs produces higher-quality surrogate labels and further improves the model from the expanded training set. Second, even though ALCATRAs achieves strong initial performance, its model improvement rate remains comparable to, or even faster than, other AL frameworks during the early stages when budget consumption is low (around less than 30\%). This is due to ALCATRAs' cost-effective task-selection policy and adaptive task latent representation learning, together with the selection of instances with highly informative task features and budget-efficient costs. We provide more detailed analysis for these two components in the following two subsections.


\subsection{Cost Efficiency of Task Acquisition Policies}

We next analyze the cost efficiency of different task-selection policies. We apply the trained policies to the same held-out test dataset and summarize the resulting task costs in Table \ref{tab:sim-costs}, comparing Streak Learning against a random selection policy, a direct-search approach \citep{zhao2015new}, and a feature acquisition baseline. The efficiency cost, which compares the estimated and optimal task-selection order, serves as the main metric for evaluating policy optimality.

\begin{table}[H]
    \centering
    \def\arraystretch{1.0}%
    \resizebox{\textwidth}{!}{
\begin{tabular}{|c|c|l|c||ccc|c||c|}
\hline
Q & Initial Labeled Size & Task-selection Policy & Task Costs & Ranking Costs & Order Costs & Num. Streaks & Efficiency Costs & Total Task Costs \\
\hline
\multirow{10}{*}{5} & \multirow{5}{*}{500} & Random Policy & 0.597 (0.002) & 0.477 (0.080) & 0.940 (0.004) & 0.250 (0.007) & 0.700 (0.038) & 0.679 (0.030) \\
 &  & Direct-Search & 0.854 (0.034) & 0.459 (0.084) & 0.907 (0.078) & 0.460 (0.385) & 0.682 (0.062) & 0.715 (0.054) \\
 &  & Feature Acquisition & 0.966 (0.041) & 0.552 (0.082) & 0.941 (0.037) & 0.294 (0.180) & 0.742 (0.051) & 0.784 (0.047) \\
 &  & \textbf{Streak-Learning} & \textbf{0.264} (0.070) & \textbf{0.144} (0.072) & \textbf{0.764} (0.121) & \textbf{0.765} (0.327) & \textbf{0.473} (0.070) & \textbf{0.431} (0.052) \\
  \cline{3-9}
 &  & \textbf{Improvement (to Best)} & \textbf{55.779\%} & \textbf{68.627\%} & \textbf{15.766\%} & \textbf{66.304\%} & \textbf{30.645\%} & \textbf{36.524\%} \\
\cline{2-9}
 & \multirow{5}{*}{1500} & Random Policy & 0.597 (0.002) & 0.477 (0.080) & 0.940 (0.004) & 0.250 (0.007) & 0.700 (0.038) & 0.679 (0.030) \\
 &  & Direct-Search & 0.869 (0.032) & 0.485 (0.080) & 0.905 (0.069) & 0.472 (0.343) & 0.693 (0.054) & 0.727 (0.047) \\
 &  & Feature Acquisition & 0.998 (0.003) & 0.583 (0.091) & 0.962 (0.035) & 0.189 (0.177) & 0.764 (0.050) & 0.810 (0.039) \\
 &  & \textbf{Streak-Learning} & \textbf{0.225} (0.059) & \textbf{0.096} (0.067) & \textbf{0.775} (0.123) & \textbf{0.700} (0.317) & \textbf{0.459} (0.075) & \textbf{0.410} (0.051) \\
  \cline{3-9}
 &  & \textbf{Improvement (to Best)} & \textbf{62.312\%} & \textbf{79.874\%} & \textbf{14.365\%} & \textbf{48.305\%} & \textbf{33.766\%} & \textbf{39.617\%} \\
\hline
\multirow{10}{*}{10} & \multirow{5}{*}{500} & Random Policy & 0.540 (0.004) & 0.540 (0.052) & 0.984 (0.001) & 0.112 (0.003) & 0.755 (0.026) & 0.712 (0.021) \\
 &  & Direct-Search & 0.591 (0.015) & 0.600 (0.053) & 0.990 (0.008) & 0.079 (0.069) & 0.785 (0.029) & 0.746 (0.023) \\
 &  & Feature Acquisition & 0.789 (0.038) & 0.569 (0.055) & 0.988 (0.010) & 0.110 (0.089) & 0.769 (0.030) & 0.773 (0.026) \\
 &  & \textbf{Streak-Learning} & \textbf{0.158} (0.051) & \textbf{0.233} (0.090) & \textbf{0.950} (0.053) & \textbf{0.282} (0.261) & \textbf{0.594} (0.056) & \textbf{0.505} (0.045) \\
  \cline{3-9}
 &  & \textbf{Improvement (to Best)} & \textbf{70.741\%} & \textbf{56.852\%} & \textbf{3.455\%} & \textbf{151.786\%} & \textbf{21.325\%} & \textbf{29.073\%} \\
\cline{2-9}
 & \multirow{5}{*}{1500} & Random Policy & 0.539 (0.004) & 0.536 (0.055) & 0.985 (0.001) & 0.112 (0.003) & 0.753 (0.027) & 0.710 (0.022) \\
 &  & Direct-Search & 0.590 (0.015) & 0.594 (0.056) & 0.989 (0.012) & 0.088 (0.095) & 0.781 (0.031) & 0.743 (0.024) \\
 &  & Feature Acquisition & 1.000 (0.000) & 0.607 (0.057) & 0.991 (0.013) & 0.089 (0.128) & 0.789 (0.032) & 0.831 (0.025) \\
 &  & \textbf{Streak-Learning} & \textbf{0.110} (0.041) & \textbf{0.172} (0.099) & \textbf{0.930} (0.064) & \textbf{0.374} (0.285) & \textbf{0.558} (0.062) & \textbf{0.469} (0.050) \\
  \cline{3-9}
 &  & \textbf{Improvement (to Best)} & \textbf{79.592\%} & \textbf{67.910\%} & \textbf{5.584\%} & \textbf{233.929\%} & \textbf{25.896\%} & \textbf{33.944\%} \\
 \hline
\end{tabular}
}

\caption{Estimated acquisition costs of task-selection policies applied to the test dataset. Policies are trained from an AL process with an unlabeled data size $N^U=1000$ and maximum budget $B=500$. Standard errors are provided in parentheses. Improvement rates compare Streak-Learning to the best performer of competing methods. The total task cost is computed as a weighted combination of task costs and efficiency costs, with each scaling factors of 0.2 and 0.8.}
\label{tab:sim-costs}
\end{table}

According to Table~\ref{tab:sim-costs},  Streak Learning achieves the lowest acquisition costs across the categories evaluated. Specifically, the task costs suggest that Streak Learning selects the fewest number of tasks, while the efficiency costs indicate that it aligns most closely with the optimal task-selection order. This demonstrates that Streak Learning not only focuses on reducing the number of selected tasks but also prioritizes selecting the most necessary tasks. In addition, when examining the number of streaks, we observe that Streak Learning maintains a substantially higher number of consecutive optimal decisions from the start of the acquisition process relative to competing methods, and significantly enhances its acquisition efficiency. Finally, we remark that the worsened performance of the direct-search and feature-acquisition methods compared to the random policy is largely due to their dependence on large sample sizes. Given that the task ordering problem is inherently of exponential complexity, effectively learning the optimal task-selection policy requires an unrealistically large amount of labeled data in the active learning setting.  This is consistent with the variance-dominance result of Theorem~\ref{thm:variance_reduction_main}, and indicates improved sample efficiency of Streak Learning over the evaluated policy-learning algorithms under the studied settings.

\subsection{Adaptive Latent Task Representation Learning Performance}

We evaluate the quality and sample efficiency of the latent task representations learned by Online-ELBO \eqref{eq: online-ELBO}, comparing against mean imputation, random forest, autoencoder, and standard VAE baselines. Performance is measured by mean-squared error (MSE) between the imputed and ground-truth task features, reported in Table \ref{tab:sim:surr}.

\begin{table}[H]
    \centering
    \def\arraystretch{1.0}%
    \resizebox{\textwidth}{!}{
       \begin{tabular}{|c|c|ccccc|cc|}
        \hline
        Q & Labeled Size & Mean & Random Forest & AutoEnoder & VAE & \textbf{O-VAE} & Oracle & \textbf{Reduction-rate} \\
        \hline
        \multirow[t]{3}{*}{5} & 150 & 62.23 (10.09) & 64.18 (10.63) & 61.78 (10.02) & 61.91 (10.08) & \textbf{61.59} (10.07) & 60.58 (10.14) & \textbf{15.64\%} \\
         & 500 & 62.12 (10.07) & 63.04 (10.35) & 61.89 (10.07) & 61.90 (10.07) & \textbf{61.57} (10.07) & 59.36 (10.08) & \textbf{12.60\%} \\
         & 1500 & 62.15 (10.07) & 62.49 (10.19) & 61.86 (10.10) & 61.94 (10.08) & \textbf{61.60} (10.07) & 59.32 (10.08) & \textbf{10.37\%} \\
         \hline
        \multirow[t]{3}{*}{10} & 150 & 71.48 (9.43) & 73.21 (9.57) & 72.72 (9.14) & 59.24 (6.81) & \textbf{59.04} (6.77) & 58.33 (6.80) & \textbf{21.87\%} \\
         & 500 & 62.59 (7.41) & 63.29 (7.56) & 64.29 (7.75) & 59.59 (6.77) & \textbf{59.25} (6.76) & 57.64 (6.78) & \textbf{17.32\%} \\
         & 1500 & 65.46 (7.75) & 65.73 (7.80) & 66.56 (8.13) & 61.05 (6.85) & \textbf{60.50} (6.81) & 57.40 (6.77) & \textbf{15.17\%} \\
         \hline
        \multirow[t]{3}{*}{15} & 150 & 72.28 (6.37) & 74.40 (6.64) & 72.45 (6.10) & 60.24 (5.52) & \textbf{59.45} (5.51) & 58.75 (5.53) & \textbf{53.12\%} \\
         & 500 & 64.90 (5.89) & 70.70 (7.13) & 66.37 (5.88) & 59.70 (5.54) & \textbf{59.32} (5.52) & 58.07 (5.52) & \textbf{23.35\%} \\
         & 1500 & 60.67 (5.65) & 68.50 (6.26) & 64.18 (5.83) & 59.84 (5.60) & \textbf{59.41} (5.57) & 58.00 (5.51) & \textbf{23.47\%} \\
        \hline
        \end{tabular}
}
    \caption{Empirical MSE scores between the learned task representations and the ground-true task features. The oracle score is obtained by applying a VAE to the fully observed task datasets. Standard errors are reported in parentheses. Reduction rates measure the reduction in MSEs from the best competing methods to the oracle score achieved by the proposed online-ELBO.}
    \label{tab:sim:surr}
\end{table}

From Table \ref{tab:sim:surr}, we observe that the proposed online variational method (O-VAE)  achieves the lowest reconstruction error (MSE) across the studied configurations. Notably, the reduction rate increases as more task features are involved and the labeled sample size decreases. This trend is expected since a larger number of available tasks increases the sparsity of the partially observed task features. Without sufficient observations for each task outcome, it is more challenging to learn task latent representation, as can be observed in the deteriorating performance of the competing imputation methods. In contrast, our method, by incorporating the dynamic task-selection probabilities into the representation learning process, is able to put more emphasis on the underrepresented features, which helps mitigate the task selection bias and further enhances the VAE framework.

\section{Data Application} \label{sec:data}

In this section we apply ALCATRAs to the UCI heart disease cohort \citep{heart_disease_45} to make two points. First, ALCATRAs improves predictive performance under a fixed acquisition budget relative to baselines. Second, and equally important for resource-constrained decision-making, the learned acquisition decisions are interpretable: they admit a transparent ranking of auxiliary information by predictive contribution per unit cost, and the ranking aligns with established cardiovascular risk-stratification practice. A parallel analysis on the FashionMNIST benchmark, in which the analogous ranking can be read from the patch-reveal order, is provided in Section~\ref{chapter3-appendix: DA-fashion} of the Supplementary Materials.

The cardiovascular screening example illustrates a common deployment asymmetry in healthcare. Expensive diagnostics may be available during model development and validation but are often unavailable during routine screening because of cost, accessibility, or workflow constraints. Building a model that generalizes from a richly-tested labeled cohort to a deployment population seen only through routine measurements is therefore not specific to the UCI cohort but is a recurring structural problem in clinical prediction.

The UCI Heart Disease dataset is a small-scale medical dataset including 303 patients who have undergone various diagnostic tests to assess their risks of heart disease. The predictive outcome is binary: either presence or absence of the disease. A total of 10 health-related variables are recorded. In this analysis, we consider easily obtainable features, such as demographics and chest-pain descriptions, as basic covariates with no acquisition costs. Meanwhile, laboratory measurements are treated as task-specific features which incur additional costs. We list the exact costs from public UCI Health's pricing data in Table~\ref{tab:da2-costs}.

\begin{table}[H]
    \centering
      \resizebox{\textwidth}{!}{
    \begin{tabular}{|c|c|c|c|c|c|c|c|c|c|c|}
    \hline
    & \multicolumn{5}{c|}{\textbf{Basic Covariates}} & \multicolumn{5}{c|}{\textbf{Task features}} \\ \hline
    Name & Age & Sex & Chest pain & Blood pressure & Heart rate &  Cholesterol & Blood sugar & EKG & Fluoroscopy & Stress test \\ \hline
    Costs & \multicolumn{5}{c|}{\$0} & \$7.2 & \$7.8 & \$67.5 & \$252.3 & \$663.6 \\ \hline
    \end{tabular}
     }
    \caption{Acquisition costs for basic covariates and task-specific features in the UCI Heart Disease dataset. Cost estimates are based on publicly available hospital pricing data.}
    \label{tab:da2-costs}
\end{table}

\begin{figure}[H]
    \centering
    \includegraphics[width=0.55\linewidth]{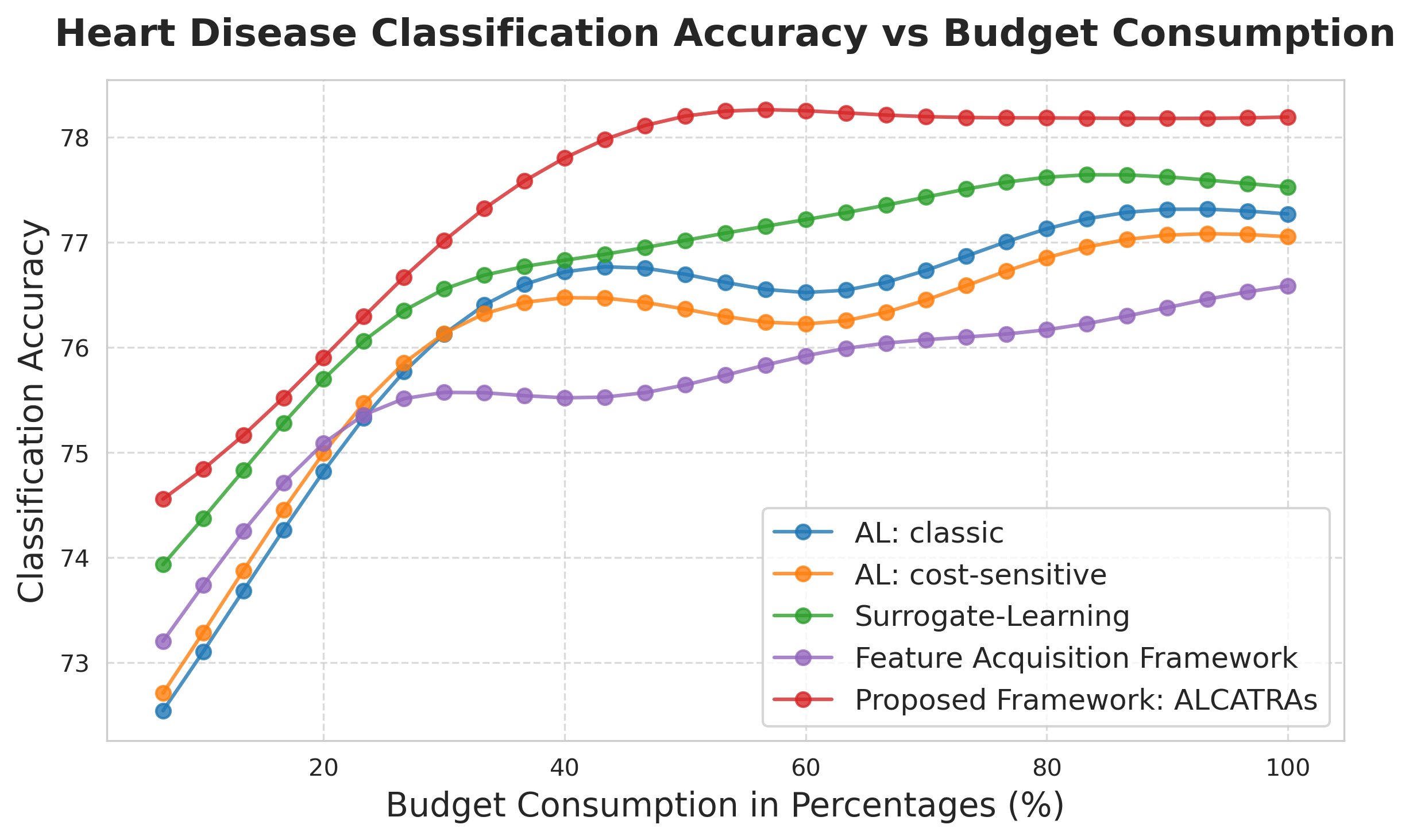}
    \caption[]{Trajectories of model predictive performance in classification accuracy versus the budget consumption, averaged over 50 experiment repetitions. Each experiment includes a random selection of 50 labeled patients, 150 unlabeled patients, and 103 patients for testing. Maximum budget are \$19755 (\$131.70 $\times$ 150), where each clinical visit costs $\$131.70$. A 100\% budget consumption indicates that all budget has been utilized.}
    \label{fig:heart-disease-res}
\end{figure}

Predictive models are trained only on the basic covariates and results averaged over 50 experiments are shown in Figure~\ref{fig:heart-disease-res}. Similar to previous findings,  ALCATRAs improves classification accuracy and sample efficiency over baselines under the studied conditions. In particular, while surrogate labels have notably improved both stability and accuracy over classical and cost-sensitive AL, ALCATRAs further enhances these advantages by leveraging informative task features to generate more accurate surrogate labels. Additionally, compared to the feature acquisition framework, ALCATRAs makes predictions solely based on basic covariates without the need to model separate belief states after each acquisition step. This highlights the efficiency of incorporating task information into surrogate labels, especially under small-sample conditions in AL.

\subsubsection*{Interpretation of learned acquisition decisions}

We also follow up the analysis by examining the interpretability of the task-selection policy. Since the relative importance of each medical diagnostic is less straightforward, we train a Random Forest model on the full dataset and extract feature importance scores, which are then inversely weighted by their corresponding diagnostic costs. For comparison, we also construct an analogous task importance score based on the learned task-selection policy. Specifically, for task $q$, we define its importance as $\sum_{t=1}^Q (Q-t) \frac{\sum \mathbb{I}(A_t=q)}{\sum \mathbb{I}(A_t \neq 0)}$, which assigns higher weights to tasks selected earlier in the decision sequence. The resulting scores are normalized across all task features and presented in Table \ref{chapter3-appendix: feature-importance}.

\begin{table}[hbt!]
    \centering
\resizebox{0.8\textwidth}{!}{
\begin{tabular}{|c|c|ccccc|}
\hline
\multicolumn{2}{|c|}{\textbf{Name}}  & Cholesterol & Blood sugar & EKG & Fluoroscopy & Stress test\\ \hline
\multicolumn{2}{|c|}{\textbf{Costs}} & \$7.2 & \$7.8 & \$67.5 & \$252.3 & \$663.6 \\ \hline
\textbf{Streak-Learning }& \textbf{Importance} & \textbf{0.465} & 0.276 & 0.087 & 0.139 & 0.033  \\ \hline
\textbf{Random Forest} & Importance & 0.083 & 0.022 & 0.092 & 0.101 & 0.213 \\ \cline{2-7}
& \begin{tabular}{@{}c@{}} \textbf{Cost-weighted} \\ \textbf{Importance}\end{tabular} & \textbf{0.517} & 0.171 & 0.286 & 0.013 & 0.013\\ \hline
\end{tabular} 
}
\caption{Task feature importance extracted from Streak-Learning policy and Random Forests.}
\label{chapter3-appendix: feature-importance}
\end{table}

According to Table \ref{chapter3-appendix: feature-importance}, if ignoring costs, stress test, the most expensive test, is identified as the most important feature by the Random Forest model. However, after accounting for task prices, both Streak-Learning and Random Forest indicate the cholesterol test as the most informative and cost-efficient for heart disease prediction. This alignment suggests that the task-policy offers comparable interpretations.

From a clinical standpoint, the prioritization is consistent with established practice in cardiovascular risk stratification: lipid screening is recommended as a first-line, low-cost diagnostic in adult cardiovascular risk assessment, while exercise stress testing and fluoroscopic imaging are typically reserved for patients with elevated pretest probability or inconclusive earlier findings \citep{arnett2019acc}. We frame this observation as supporting interpretability rather than independent clinical validation, since the dataset is small and the comparison is across only five diagnostics.

Beyond predictive accuracy, this interpretability provides a second use case for the framework: as a decision-support tool for designing cost-effective data-collection strategies in clinical research. The learned acquisition decisions concentrate budget on diagnostics whose marginal predictive contribution is large relative to their acquisition cost, suggesting that ALCATRAs can inform protocol design, for example in pragmatic trials or registry studies, when investigators must decide which optional measurements to collect.

\section{Discussion} \label{sec:discussion}

From a statistical perspective, the framework developed in this paper addresses a recurring challenge in modern data collection: how to allocate limited measurement resources during training when some of those measurements will not be available at deployment. ALCATRAs treats the choice of which auxiliary variables to acquire and the choice of how to transfer the acquired information to the deployment-time predictor as a single learning problem under a fixed budget. The approach extends ideas from semi-supervised learning, two-phase sampling, and missing-by-design observation to settings in which the observation mechanism itself is part of the learning procedure.

Methodologically, ALCATRAs combines a sample-efficient acquisition decision rule, Streak Learning, which adapts off-policy evaluation to the small-labeled-data regime of active learning, with a policy-adjusted variational objective, Online-ELBO, which corrects the selection bias introduced when surrogate models are trained on adaptively acquired data. The supporting theory establishes a localized excess risk bound for the target predictor, finite-sample efficiency of the acquisition rule relative to standard importance weighting, and causal identification of the variational objective. Together these guarantees show that selectively acquired auxiliary information can improve prediction efficiency under budget constraints rather than merely improving raw accuracy.

Two practical scope conditions are worth noting. First, the gains of the framework depend on the auxiliary variables being sufficiently informative for the target prediction; under-informative auxiliaries reduce ALCATRAs to standard cost-sensitive active learning. Second, acquisition costs are assumed known and fixed; cost-estimation error and budget uncertainty are outside the present scope.

Several directions are natural extensions of the present work. First, the shared-budget assumption can be relaxed to allow individual-level budget constraints, with acquisition decisions made jointly across subjects. Second, the surrogate objective can be tightened with adaptive thresholding to retain only high-confidence completions, and extended to settings in which the queried labels are themselves uncertain, as with human-feedback annotation where labels exhibit annotator-specific heterogeneity. Third, the framework extends naturally to modern multi-agent and large-language-model settings in which the auxiliary variables take the form of costly calls to specialist tools or to a multi-step reasoning chain: the acquisition decision rule selects which tool or chain step to invoke under a budget, and the policy-adjusted variational objective transfers information from invoked calls to inputs for which the calls were not made. This direction is relevant for cost-aware deployment of language-model agents where each tool call carries an inference cost.

\section{Disclosure Statement}
\label{disclosure-statement}

The authors declare that there are no conflicts of interest.

\section{Data Availability Statement}
\label{data-availability-statement}

The UCI Heart Disease dataset analyzed in this study is publicly available from the UCI Machine Learning Repository at \url{https://archive.ics.uci.edu/dataset/45/heart+disease}. The FashionMNIST benchmark used in the supplementary analysis is publicly available at \url{https://github.com/zalandoresearch/fashion-mnist}.

\phantomsection\label{supplementary-material}
\bigskip

\begin{center}

{\large\bf SUPPLEMENTARY MATERIAL}

\end{center}

\begin{description}
\item[Supplementary Material:]
An online supplement to this article is available, which contains detailed proofs of all theorems and corollaries, additional regularity assumptions, implementation details for the ALCATRAs framework, and a parallel application on the FashionMNIST benchmark (PDF file).
\end{description}

{
\small
\bibliography{reference}
}

\appendix

\end{document}


\def\spacingset#1{\renewcommand{\baselinestretch}%
{#1}\small\normalsize} \spacingset{1.8}

\begin{center}
{\large\bf Supplementary Materials \\
to Multi-Task Active Learning with Efficient Resource Allocation}
\end{center}
\tableofcontents
\newpage

\section{Derivation of Streak Learning estimator} \label{chapter3-appendix: sec1}

In this appendix we derive the following estimator for the Streak Learning in Section 2.3:
\begin{equation}
\text{VF}^{\text{Streak}}(\pi_1, \ldots,\pi_{Q})  =\mathbb{E} \left\{ \sum_{\delta=1}^{\bar{T}+1} \phi(\delta) \cdot \frac{R \cdot {\prod_{t=1}^{\delta-1}} \;\mathbb{I}(\pi_t(H_t)=A_t) \cdot {\sum_{t=\delta}^{\bar{T}}} \;\mathbb{I}(\pi_t(H_t)=A_t)))}{\prod_{t=1}^T \pi^0_t(A_t | H_t)}\right\}. \notag
\end{equation}

\begin{proof}
First, we assume the following three common causal assumptions:
\begin{assumption}
(Stable Unit Treatment Value Assumption, SUTVA \citep{rubin1980randomization}) Potential outcomes of each subject are not affected by the treatments assigned to other subjects (no interference); there is no different form of each treatment level (no hidden variations).
\label{A-SUTVA}
\end{assumption}
\vspace{-1em}
\begin{assumption}
(Sequential Ignorability \citep{robins1986new}) Treatment assignments are independent of potential future outcomes, conditional on the history up to the current time $j$, i.e., $R_{j:T} \perp A_j  \;|\;H_j$.
\label{A-ignorability}
\end{assumption}
\vspace{-1em}
\begin{assumption}
(Positivity) Any subjects are possibly assigned for all treatments, i.e., $P(A_j=a_j \;|\;H_j) > 0$ for any $a_j \in \mathcal{A}$ and $1 \le t \le T$.
\label{A-pos}
\end{assumption}

Under the above three assumptions, standard by Radon-Nikodym theorem, the potential outcome of the task-selection policy $\bm{\pi} = \{\pi_t\}_{t=1}^T$ can be written as an inverse-propensity weighted form of the observed reward $R$,
\begin{equation}
    \mathbb{E}^{\bm{\pi}}\{R\} = \mathbb{E}\left\{ R \cdot \frac{\prod_{t=1}^T \mathbb{I}(A_t=\pi_t(H_t)}{\prod_{t=1}^T \pi_t^0(A_t | H_t)}\right\} \label{chapter3-app: IPWE}
\end{equation}
However, as not every task has to be selected and queried, the number of decision stages could be different among individuals. To account for this variability in imbalanced trajectories, we adopt the auxiliary approach introduced in \cite{goldberg2012q}. Specifically, we consider a random variable $\Tilde{T}$, indicating the decision-making process stops at time stage $\Tilde{T}$, $1 \le \Tilde{T} \le T$, such that no further action at $\Tilde{T} < t \le T$ would be made. Then the auxiliary states and actions after stage $\Tilde{T}$ are defined as null set and uniform draws from the action space $\mathcal{A}$ respectively, i.e., $\{H_t'=\emptyset, a_t^{'} \sim \text{Unif} (\mathcal{A})\}_{t=\Tilde{T}+1}^{T}$. As a result, the likelihood of a trajectory under the behavioral policy $\pi^0$ has the following form, 
$$
f_1(H_1)\pi_t^0 (A_1|H_1) \prod_{t=2}^{\Tilde{T}}\{f_t(H_t|H_{t-1}) \pi^0_t(A_t|H_t)\} \prod_{t=\Tilde{T}+1}^T \{1 \cdot \pi^0_t(A_t^{'}|H_t^{'})\}.
 $$

In addition, we let the policy decision on any auxiliary state is selected as a fixed action, i.e., $\pi_t(H_t^{'} = \emptyset)= a_0 \in \mathcal{A}$. Then, the likelihood of a trajectory under the policy $\bm{\pi}$ has the form, 
$$
f_1(H_1)\mathbb{I}(\pi_1(H_1)=A_1) \prod_{t=2}^{\Tilde{T}}\{f_t(H_t|H_{t-1}) \mathbb{I}(\pi_t(H_t)=A_t)\} \prod_{t=\Tilde{T}+1}^T \{1 \cdot  \mathbb{I}(\pi_t(H_t^{'})=A_t^{'})\}\}.
 $$

Applying Radon-Nikodym theorem again, Equation \eqref{chapter3-app: IPWE} can be written as a aggregated potential outcome under each specified trajectory length $\Tilde{T}$, i.e.,
\begin{equation}
\mathbb{E}\left\{ R \cdot \frac{\prod_{t=1}^T \mathbb{I}(A_t=\pi_t(H_t))}{\prod_{t=1}^T \pi_0(A_t | H_t)}\right\} = \mathbb{E}_{\Tilde{T}}  \left\{ R \cdot \frac{\prod_{t=1}^{\Tilde{T}} \mathbb{I}(A_t=\pi_t(H_t)) \cdot \prod_{t=\Tilde{T}+1}^{T}  \mathbb{I}(A_t^{'}=\pi_t(H_t^{'}))}{\prod_{t=1}^{\Tilde{T}} \pi_t^0(A_t | H_t) \cdot \prod_{t=\Tilde{T}+1}^T \pi^0_t(A_t^{'}|H_t^{'})} \right\} 
\label{chapter3-app: IPWE-TildeT}
\end{equation}

Equation \eqref{chapter3-app: IPWE-TildeT} accommodates different number of decision stages, but still has the form of inverse-propensity weighting, which lacks sample efficiency due to \textit{curse of full-matching} introduced in \cite{ye2024stage}. While their $k$-IPWE and SAL solutions improve the efficiency by allowing observed actions not strictly aligned with the policy decisions, such relaxation of misalignment can takes place at any decision stage. In resource allocation, however, one want to keep making optimal decisions unless a mistake has been unfortunately made. For Streak Learning, we leverage the sample efficiency of $k$-IPWE while also accounting for the unique decision structure inherent to resource allocation problems. Specifically, we introduce a random variable $\delta$, $1 \le \delta \le \Tilde{T}$, to indicate the first decision stage where a mistake happens. Then Equation \eqref{chapter3-app: IPWE-TildeT} can be further decomposed into,
\begin{equation}
\mathbb{E}_{\Tilde{T}}  \left\{ R \cdot \frac{\prod_{t=1}^{\Tilde{T}} z_t \cdot \prod_{t=\Tilde{T}+1}^{T} z_t^{'}}{\prod_{t=1}^{\Tilde{T}} \pi_t^0 \cdot \prod_{t=\Tilde{T}+1}^T \pi^{0'}_t} \right\} 
= \mathbb{E}_{\Tilde{T}, \delta}  \left\{ R \cdot \frac{\prod_{t=1}^{\delta-1} z_t \cdot \prod_{t=\delta}^{\Tilde{T}} z_t \cdot  \prod_{t=\Tilde{T}+1}^{T} z^{'}_t}{\prod_{t=1}^{\Tilde{T}} \pi_t^0\cdot \prod_{t=\Tilde{T}+1}^T \pi^{0'}_t} \right\},
\label{chapter3-app: IPWE-delta}
\end{equation}
where $z_t \doteq \mathbb{I}(A_t= \pi_t(H_t))$, $z_t^{'} \doteq \mathbb{I}(A_t= \pi_t(H^{'}_t))$, $\pi^0$ short for $\pi^0_t(A_t|H_t)$, and $\pi^{0'}_t$ short for $\pi^0_t(A^{'}_t | H^{'}_t)$ for ease of denotations. Since $\delta$ is the first stage where decision misalignment takes place, we adopt the $k$-IPWE approach by assuming there exists $K$ number of decision stages, $1 \le K \le \Tilde{T}-\delta+1$, whose observed actions are aligned with the policy decisions among the stages $\delta \le t \le \Tilde{T}$. Assume among the $K$ number of stages, stages with index $\mathcal{K}$ have $z_{t \in \mathcal{K}} = 1$ while $z_{t \notin \mathcal{K}} = 0$. By relaxing the decision misalignments, we approximate estimator \eqref{chapter3-app: IPWE-delta} with the following,
\begin{align}
& \mathbb{E}_{\Tilde{T}, \delta, K}  \left\{ R \cdot \frac{\prod_{t=1}^{\delta-1} z_t \cdot \prod_{t \in \mathcal{K}}^{\Tilde{T}} z_t \cdot \prod_{t \notin \mathcal{K}} (1 - z_t) \prod_{t=\Tilde{T}+1}^{T} z^{'}_t}{\prod_{t=1}^{\Tilde{T}} \pi_t^0\cdot \prod_{t=\Tilde{T}+1}^T \pi^{0'}_t} \right\} \notag \\
&= \mathbb{E}_{\Tilde{T}, \delta, K} \left\{ { \Tilde{T}-\delta+1 \choose K}^{-1} \sum_{\mathcal{K} }R \cdot \frac{\prod_{t=1}^{\delta-1} z_t \cdot \prod_{t \in \mathcal{K}}^{\Tilde{T}} z_t \cdot \prod_{t \notin \mathcal{K}} (1 - z_t) \prod_{t=\Tilde{T}+1}^{T} z^{'}_t}{\prod_{t=1}^{\Tilde{T}} \pi_t^0\cdot \prod_{t=\Tilde{T}+1}^T \pi^{0'}_t} \right\} \notag \\
&= \mathbb{E}_{\Tilde{T}, \delta, K} \left\{ c_0 \cdot R \cdot \frac{\prod_{t=1}^{\delta-1} z_t \cdot \mathbb{I}(\sum_{t=\delta}^{\Tilde{T}} z_t = K) \cdot  \prod_{t=\Tilde{T}+1}^{T} z^{'}_t}{\prod_{t=1}^{\Tilde{T}} \pi_t^0\cdot \prod_{t=\Tilde{T}+1}^T \pi^{0'}_t} \right\} \label{chapter3-appendix: intersect},
\end{align}
where $c_0 = {\Tilde{T}-\delta + 1 \choose K}^{-1}$ and Equation \eqref{chapter3-appendix: intersect} is due to only one combination from stages $\delta$ to $\Tilde{T}$ having the observed action $K$-partially matched with the policy decisions. Lastly, assuming the random $K$ matching number increases linearly, we obtain the value function for Streak Learning by combining Equation \eqref{chapter3-appendix: intersect} with each possible $K$ number.
\begin{align}
&\text{VF}^{\text{Streak}} = \sum_{k=1}^{\Tilde{T}-\delta+1} \frac{k}{\Tilde{T}-\delta+1} \cdot \mathbb{E}_{\Tilde{T}, \delta, K=k} \left\{ R \cdot \frac{\prod_{t=1}^{\delta-1} z_t \cdot \mathbb{I}(\sum_{t=\delta}^{\Tilde{T}} z_t = K) \cdot  \prod_{t=\Tilde{T}+1}^{T} z^{'}_t}{\prod_{t=1}^{\Tilde{T}} \pi_t^0\cdot \prod_{t=\Tilde{T}+1}^T \pi^{0'}_t} \right\} \notag \\
&= \sum_{k=1}^{\Tilde{T}-\delta+1}  \mathbb{E}_{\Tilde{T}, \delta, K=k} \left\{ R \cdot \frac{\prod_{t=1}^{\delta-1} z_t \cdot \left( \frac{k}{\Tilde{T}-\delta+1} \cdot \mathbb{I}(\sum_{t=\delta}^{\Tilde{T}} z_t = k)\right) \cdot  \prod_{t=\Tilde{T}+1}^{T} z^{'}_t}{\prod_{t=1}^{\Tilde{T}} \pi_t^0\cdot \prod_{t=\Tilde{T}+1}^T \pi^{0'}_t}  \;\middle|\; K=k\right\} \cdot P(k)\notag \\
&= \sum_{k=1}^{\Tilde{T}-\delta+1}  \mathbb{E}_{\Tilde{T}, \delta, K=k} \left\{ R \cdot \frac{\prod_{t=1}^{\delta-1} z_t \cdot \left(\frac{1}{\Tilde{T}-\delta+1} \cdot \sum_{t=\delta}^{\Tilde{T}} z_t \right) \cdot  \prod_{t=\Tilde{T}+1}^{T} z^{'}_t}{\prod_{t=1}^{\Tilde{T}} \pi_t^0\cdot \prod_{t=\Tilde{T}+1}^T \pi^{0'}_t}  \;\middle|\; K=k\right\} \cdot P(K=k) \label{chapter3-appendix:VF-1} \\
&= \mathbb{E}_{\Tilde{T}, \delta} \left\{ R \cdot \frac{\prod_{t=1}^{\delta-1} z_t \cdot \left(\frac{1}{\Tilde{T}-\delta+1} \cdot \sum_{t=\delta}^{\Tilde{T}} z_t \right) \cdot  \prod_{t=\Tilde{T}+1}^{T} z^{'}_t}{\prod_{t=1}^{\Tilde{T}} \pi_t^0\cdot \prod_{t=\Tilde{T}+1}^T \pi^{0'}_t} \right\},  \label{chapter3-appendix:VF-2}
\end{align}

where Equation \eqref{chapter3-appendix:VF-1} is because $k \cdot \mathbb{I}(\sum_{t=\delta}^{\Tilde{T}} z_t = k) = \sum_{t=\delta}^{\Tilde{T}} z_t$ in the event when $K=k$. Lastly, suppose the weighting scheme for the streaks is $\phi(\delta)$, we obtain the final value function from the last result \eqref{chapter3-appendix:VF-2} as desired,
\begin{equation}
\text{VF}^{\text{Streak}} =  \mathbb{E}_{\Tilde{T}} \left\{  \sum_{\delta=1}^{\Tilde{T}} \phi(\delta) \cdot R \cdot \frac{\prod_{t=1}^{\delta-1} z_t \cdot \frac{1}{\Tilde{T}-\delta+1} \cdot \sum_{t=\delta}^{\Tilde{T}} z_t \cdot  \prod_{t=\Tilde{T}+1}^{T} z^{'}_t}{\prod_{t=1}^{\Tilde{T}} \pi_t^0\cdot \prod_{t=\Tilde{T}+1}^T \pi^{0'}_t} \right\}.
\end{equation}
We remark that in the main text, we write $z^{'}_t$ and $\pi^{0'}_t$ as $z_t$ and $\pi^0_t$ for readability. But once the decision process terminates, the observed actions and policy decisions are selected according to the auxiliary setup.
\end{proof}

\section{Theoretical proof}
\subsection{Theorem 1: Localized Excess Risk for Surrogate Learning}
Let $R(f) = \mathbb{E}\{(f(X) - y)^2\}$ and $\hat{R}_L(f) = \frac{1}{N^L} \sum_{i=1}^{N^L}(f(X_i) - y_i)^2$ respectively denote the population and empirical risk on the labeled set for a target predictive function $f \in \mathcal{F}: \mathcal{X} \mapsto \mathcal{Y}$. Similarly, for a surrogate function $g \in \mathcal{G}: \mathcal{X} \times \mathcal{W} \mapsto \mathcal{Y}$, we define the corresponding risk functions as $R(g) = \mathbb{E}\{(g(X, W) - y)^2\}$ and $\hat{R}_L(g) = \frac{1}{N^L} \sum_{i=1}^{N^L} (g(X_i, W_i) - y_i)^2$.

Our goal is to bound the excess risks of the target predictive model: 
\begin{equation}
R(\hat{f}) - R(f^*) = R(\hat{f}) - \hat{l}_N(\hat{f}) + \underbrace{\hat{l}_N(\hat{f}) - \hat{l}_N(f^*)}_{\le 0} + \hat{l}_N(f^*) - R(f^*), \label{eq: sup-f}
\end{equation}
where $f^* = \argmin_{f \in \mathcal{F}} R(f)$ is the ground-true predictive function and $\hat{f} =\argmin_{f \in \mathcal{F}} \hat{l}_N(f)$ is the estimator minimizing the empirical surrogate learning loss: $\hat{l}_N(f; g) = \hat{R}_L(f) + \|f-g\|^2_U$. The second term, $\|f-g\|^2_U = \frac{1}{N^U} \sum_{i=1}^{N^U} (f(X_i) - g(X_i, \hat{W}_i))^2$, measures the target function risks to the surrogate labels obtained from a given surrogate model $g$ on the unlabeled set.

To analyze the generalization properties with the addition of auxiliary task information, we consider a restricted function class $\mathcal{F}_\epsilon$. This class contains all target functions have an empirical discrepancy of at most $\epsilon$ compared to the surrogate $\Tilde{g}$ on the unlabeled set:
\begin{equation}
    \mathcal{F}_\epsilon \coloneqq \{f \in \mathcal{F}: \|f(X) - \Tilde{g}(X)\|_U \le \epsilon\}.
\end{equation}
Here, $\tilde{g}(X) = \hat{g}(X, \hat{W}(X)):  \mathcal{X} \mapsto \mathcal{Y}$ is the online surrogate model and $\hat{g}$ minimizes the empirical risks $\hat{R}_L(g)$. Under the assumption that both the best estimator $\hat{f}$ and the true target function $f^*$ reside within this restricted class (i.e., $\hat{f}, f^* \in \mathcal{F}_\epsilon$), the risk bounds in Equation \eqref{eq: sup-f} can be further up-bounded by the following:
\begin{align}
     R(\hat{f}) - R(f^*) &\le 2 \sup_{f \in \mathcal{F}_\epsilon} |R(f) - \hat{l}_N(f)| = 2 \sup_{f \in \mathcal{F}_\epsilon} \left| R(f) - \left( \hat{R}_L(f) + \|f - \hat{g}\|_U^2 \right) \right| \notag \\
     &\le 2 \sup_{f \in \mathcal{F}_\epsilon} |R(f) - \hat{R}_L(f)| + 2\epsilon^2, \label{eq:risk_limit_jasa}
\end{align}

In the following, we establish a uniform convergence bound for the excess risk over the restricted class $\mathcal{F}_\epsilon$. We rely on the concentration results established by \citet{bartlett2005local}.

\subsubsection{Lemma 1 (Bartlett's Localized Concentration Lemma)}\label{sec:bartlett_lemma}
\begin{lemma}
(Adapted from Theorem 2.1 in \citet{bartlett2005local}). \textit{Let $\mathcal{F}$ be a class of functions that map $\mathcal{X}$ into $[a,b]$. Assume that there is some $r>0$ such that for every $f \in \mathcal{F}$, $\text{Var}(f(X_i)) \le r$. Then, for every $x > 0$, with probability at least $1 - e^{-x}$,}
\begin{equation}
\sup_{f \in \mathcal{F}}(P_n f - P f) \le \inf_{\alpha > 0} \left( 2(1+\alpha)\mathbb{E}\mathscr{R}_n (\mathcal{F}) + \sqrt{\frac{2rx}{n}} + (b-a)\cdot \left( \frac{1}{3} + \frac{1}{\alpha}\right) \cdot \frac{x}{n} \right), \notag
\end{equation}
where $\alpha$ controls the star-hull of $\mathcal{F}$ around the reference function $f_0$, i.e., $\text{star}(\mathcal{F}, f_0)=\{f_0+\alpha(f-f_0): f \in \mathcal{F}, \alpha \in [0,1]\}$
\label{lemma:bartlett}
\end{lemma}
This result directly relates the convergence rate of the empirical process to the variance of the function class. Such a formulation is well-suited to our problem as we expect high-quality surrogate learning can reduce the variance of the localized function space $\mathcal{F}_\epsilon$, which in turn tightens the resulting generalization bounds. To apply this concentration result, we first establish the following regularity assumptions.

\begin{assumption}[Boundedness]\label{ass:boundedness}
We assume that the labeling targets, target model predictions, and surrogate predictions are strictly bounded almost surely such that $|y| \le B$ and $|f(X)| \le B$ for all $f \in \mathcal{F}$, and $|g(X, W)| \le B$ for all $g \in \mathcal{G}$ under a bounding constant $B > 0$.
\end{assumption}

\begin{assumption}[Metric Entropy Growth]\label{ass:entropy}
Let $\mathcal{N}(\delta, \mathcal{H}, \|\cdot\|_n)$ denote the $\delta$-covering number of a function class $\mathcal{H}$ with respect to the empirical $L_2$ norm on a sample with size $n$. We assume that the log-covering numbers (metric entropy) of the target class $\mathcal{F}$, surrogate class $\mathcal{G}$, and variational autoencoder class $\mathcal{Q}$ satisfy the following polynomial growth boundaries:
\begin{equation}
    \log \mathcal{N}(\delta, \mathcal{F}, \|\cdot\|_n) \le C_{\mathcal{F}} \left(\frac{1}{\delta}\right)^{\alpha_{\mathcal{F}}}, \log \mathcal{N}(\delta, \mathcal{G}, \|\cdot\|_n) \le C_{\mathcal{G}} \left(\frac{1}{\delta}\right)^{\alpha_{\mathcal{G}}}, \log \mathcal{N}(\delta, \mathcal{Q}, \|\cdot\|_n) \le C_{\mathcal{Q}} \left(\frac{1}{\delta}\right)^{\alpha_{\mathcal{Q}}}
\end{equation}
for constants $C_{\mathcal{F}}, C_{\mathcal{G}}, C_{\mathcal{Q}} > 0$ and metric capacity exponents $\alpha_{\mathcal{F}}, \alpha_{\mathcal{G}},\alpha_{\mathcal{Q}} \in (0, 2)$.
\end{assumption}

\begin{assumption}[Uniform Metric Stability]\label{ass:stability}
For each sample size $n$, there exist constants $c_1', c_2', c_1'', c_2'' > 0$ such that, the uniform metric equivalence (stability) holds almost surely over both the target function class $\mathcal{F}$ and the surrogate function class $\mathcal{G}$,
\begin{align}
    c_1' \|f_1 - f_2\|_n &\le \|f_1 - f_2\|_2 \le c_2' \|f_1 - f_2\|_n \quad \forall f_1, f_2 \in \mathcal{F}, \\
    c_1'' \|g_1 - g_2\|_n &\le \|g_1 - g_2\|_2 \le c_2'' \|g_1 - g_2\|_n \quad \forall g_1, g_2 \in \mathcal{G},
\end{align}
\end{assumption}

\begin{assumption}[Surrogate Lipschitz Continuity]\label{ass:lipschitz}
Functions in the surrogate class $\mathcal{G}$ are Lipschitz continuous with respect to their task outcome variables:
\begin{equation}
 \|g(X, W) - g(X, W')\|_2 \le L_g \|W - W'\|_2 \quad \forall g \in \mathcal{G},
\end{equation}
for all $X \in \mathcal{X}$ and task features $W, W'$, under a Lipschitz constant $L_g > 0$.
\end{assumption}

\begin{assumption}[Convexity]\label{ass:convexity}
The target function class $\mathcal{F}$ is a convex set, and the completed online surrogate predictor $\tilde{g}(X) = \hat{g}(X, \hat{W}(X))$ resides within $\mathcal{F}$.
\end{assumption}

\begin{assumption}[Localized Target Realizability]\label{ass:realizability}
Both the population optimal target predictor $f^*$ and constrained empirical estimator $\hat{f}$ reside in the restricted localized class $f^*, \hat{f} \in \mathcal{F}_\epsilon \coloneqq \{ f \in \mathcal{F} : \|f - \tilde{g}\|_U \le \epsilon \}$, where $\tilde{g}(X) = \hat{g}(X, \hat{W}(X)):  \mathcal{X} \mapsto \mathcal{Y}$ is the online surrogate model.
\end{assumption}

Lemma \ref{lemma:bartlett} is applied to the loss class $\mathcal{H} = \{ l_f : f \in \mathcal{F}_\epsilon \}$ where $l_f(x,y) = (f(x) - y)^2$. After showing the Rademacher complexity of the loss class $\mathbb{E}\mathscr{R}_n (\mathcal{\mathcal{H}})$ and the upper bound of its variance $Var(l(X))$, we present the following theorem.

\begin{theorem} (Localized Excess Risk Bound for Surrogate Learning)\label{thm:main}
By Assumptions \ref{ass:boundedness}--\ref{ass:realizability}, for every $x > 0$, with probability at least $1 - e^{-x}$, the localized excess risk of the target predictive model is bounded by:
\begin{align}
    R(\hat{f}) - R(f^*) &\le  C_1 \left( \frac{\epsilon^{1-\alpha_{\mathcal{F}}/2}}{\sqrt{N^L}} \right) + 8B \sqrt{x + \ln(6)} \left( \frac{\epsilon + \sqrt{2} L_g \|W - \hat{W}\|_L + \xi}{\sqrt{N^L}} \right. \notag \\
    &\quad \left. + \sqrt{C_2} \frac{\epsilon^{1/2 - \alpha_{\mathcal{F}}/4}}{(N^U)^{1/4} \sqrt{N^L}} + \frac{\sqrt{C_3} \text{dim}(W)^{1/4} + \sqrt{C_4}}{(N^L)^{3/4}} \right) +  C_5 \frac{x + \ln(6)}{N^L} + 2\epsilon^2,
\end{align}
where $\|W - \hat{W}\|_L$ is the observed VAE reconstruction error, $\xi = \|g^* - y\|_2$ is the optimal surrogate error, and $\text{dim}(W)$ is the task outcomes dimension. Here, $C_1, C_2, C_3, C_4, C_5 > 0$ are finite capacity and tail constants depending only on the metric entropy capacity, Lipschitz outcomes continuity, and boundedness parameters of the respective function classes.
\end{theorem}

\begin{proof}[Proof of Theorem~\ref{thm:main}]
First, by the Boundedness Assumption \ref{ass:boundedness}, the loss Functions in $\mathcal{H}$ map to $[0, 4B^2]$, so the range is bounded by $b-a = 4B^2$. Then by the variance upper-bound shown later in Lemma \ref{lemma:variance}, for any $x > 0$, with probability at least $1 - e^{-x}$, the variance of $l_f$ for any $f \in \mathcal{F}_\epsilon$ is upper bounded by $r(x)$ in terms of representation errors. Applying Lemma~\ref{lemma:bartlett} to $\mathcal{H}$ on the labeled set $L$ of size $N^L$ with the variance bound $r(x)$ and choosing $\alpha = 1$, we have that with probability at least $1 - e^{-x}$:
\begin{equation}
    \sup_{f \in \mathcal{F}_\epsilon} ( \hat{R}_L(f) - R(f) ) \le 4 \mathbb{E}\mathscr{R}_{N^L}(\mathcal{L} \circ \mathcal{F}_\epsilon) + \sqrt{\frac{2 r(x) x}{N^L}} + \frac{16 B^2}{3} \frac{x}{N^L}. \label{eq:bartlett-applied}
\end{equation}
The rest of the proof focus on deriving the bounds for $\mathbb{E}\mathscr{R}_{N^L}(\mathcal{L} \circ \mathcal{F}_\epsilon)$ and $r(x)$.

\textbf{Part 1: Rademacher complexity of $\mathcal{F}_\epsilon$ on labeled sample} To relate the Rademacher complexity of the loss class to the restricted target class, we use the Lipschitz property of risks proved in Lemma \ref{lemma:lipschitz} and obtain:
\begin{equation*}
    \mathbb{E}\mathscr{R}_{N^L}(\mathcal{L} \circ \mathcal{F}_\epsilon) \le 4B \cdot \mathbb{E}\mathscr{R}_{N^L}(\mathcal{F}_\epsilon).
\end{equation*}
Then, by the standard symmetrization theorem for empirical processes $\mathbb{G}_{N^L}(f-\tilde{g})=\sqrt{N^L}(P_{N^L}-P)(f-\tilde{g})$ of the centered function class $\{f-\tilde{g}:f\in\mathcal{F}_\epsilon\}$, we obtain:
\begin{align}
    \mathbb{E}\mathscr{R}_{N^L}(\mathcal{F}_\epsilon) &\le \frac{2}{\sqrt{N^L}} \mathbb{E}\left[\sup _{f\in \mathcal{F}_\epsilon}|\mathbb{G}_{N^L}(f - \tilde{g})|\right], \label{eq:symmetrization}
\end{align}
Applying Theorem 2.14.1 in \citet{van1996weak} for sub-Gaussian empirical processes, the expected supremum is bounded by the entropy integral evaluated under the population $L_2(P)$ metric space $(\mathcal{F}_\epsilon, \|\cdot\|_2)$ up to the population radius:
\begin{align}
    \mathbb{E}\left\{\sup _{f\in \mathcal{F}_\epsilon}|\mathbb{G}_{N^L}(f - \tilde{g})|\right\} &\le C \cdot \int _{0}^{\sup_{f\in\mathcal{F}_\epsilon}
    \|f-\tilde{g}\|_2 }\sqrt{\log \mathcal{N}(\delta ,\mathcal{F}_\epsilon, \|\cdot\|_2)}\,d\delta, \label{eq:dudley-chaining}
\end{align}
By monotonicity of covering numbers and Uniform Metric Stability (Assumption \ref{ass:stability}):
\begin{equation}
    \log \mathcal{N}(\delta ,\mathcal{F}_\epsilon, \|\cdot\|_2) \le \log \mathcal{N}(\delta ,\mathcal{F}, \|\cdot\|_2) \le \log \mathcal{N}((c_2')^{-1}\delta ,\mathcal{F}, \|\cdot\|_L) \le C_{\mathcal{F}} (\delta/c_2')^{-\alpha_{\mathcal{F}}}, \label{eq:metric-stability-entropy-transition}
\end{equation}
where the second inequality can be proved by letting $\{h_1,...,h_{\mathcal{N}((c_2')^{-1}\delta ,\mathcal{F}, \|\cdot\|_L)}\} \subseteq \mathcal{F}$ be the minimal $(c_2')^{-1}\delta$-cover of the $\mathcal{F}$ space under $\|\cdot\|_{L}$ metric. Then $\|f-h\|_2 \le c_2^{'} \cdot \|f-h\|_L \le c_2^{'} \cdot (c_2^{'})^{-1} \cdot \delta \le \delta$ shows that the $h$ set also satisfies a $\delta$-cover of $\mathcal{F}$ under the population $\|\cdot\|_2$ and the covering number being the minimal set leads to the results. 

Lastly, evaluating the Dudley population integral \eqref{eq:dudley-chaining} using this metric boundary and bounding the population radius by $D_{\text{pop}} \coloneqq \sup_{f\in\mathcal{F}_\epsilon} \|f-\tilde{g}\|_2 \le c_2' \|f-\tilde{g}\|_U \le c_2' \epsilon$ under the population metric stability of Assumption \ref{ass:stability}, we obtain:
\begin{align}
    \mathbb{E}\left\{\sup _{f\in \mathcal{F}_\epsilon}|\mathbb{G}_{N^L}(f - \tilde{g})|\right\} &\le C \sqrt{C_{\mathcal{F}}} (c_2')^{\alpha_{\mathcal{F}}/2} \int_0^{c_2' \epsilon} \delta^{-\alpha_{\mathcal{F}}/2} \, d\delta =\frac{2 C \sqrt{C_{\mathcal{F}}} (c_2')^{1+\alpha_{\mathcal{F}}/2}}{2 - \alpha_{\mathcal{F}}} \epsilon^{1-\alpha_{\mathcal{F}}/2} \label{eq:expected-sup-resolved}
\end{align}
Substituting \eqref{eq:expected-sup-resolved} back into \eqref{eq:symmetrization} yields the final resolved expected Rademacher complexity:
\begin{align}
    \mathbb{E}\mathscr{R}_{N^L}(\mathcal{F}_\epsilon) &\le \frac{2}{\sqrt{N^L}} \frac{2 C \sqrt{C_{\mathcal{F}}} (c_2')^{1+\alpha_{\mathcal{F}}/2}}{2 - \alpha_{\mathcal{F}}} \epsilon^{1-\alpha_{\mathcal{F}}/2}  \le C_1 \frac{\epsilon^{1-\alpha_{\mathcal{F}}/2}}{\sqrt{N^L}}, \label{eq:rad-l-resolved-final}
\end{align}
where $C_1 = \frac{4 C \sqrt{C_{\mathcal{F}}} (c_2')^{1+\alpha_{\mathcal{F}}/2}}{2 - \alpha_{\mathcal{F}}}$ is the resolved universal localized capacity constant, and this completes the proof for the Rademacher complexity of the restricted function class.

\vspace{1em}
\textbf{Part 2: Localized Variance Bound} We bound the variance of the loss function $l_f(X,Y) = (f(X) - Y)^2$ over the localized class $f \in \mathcal{F}_\epsilon$ by its second moment, i.e., 
\begin{align}
    \text{Var}(l_f) &\le \mathbb{E}\{l_f^2\} = \mathbb{E}\{(f(X) - Y)^4\} \label{eq:var-to-risk} \\
    &\le 4B^2  \mathbb{E}\{(f(X) - Y)^2\} \quad \text{Assumption 1 Boundedness} \\
    &\le 8B^2 \left\{\underbrace{\|f(X) - \tilde{g}(X)\|^2_2}_{\text{Target-Surrogate Discrepancy}} + \underbrace{\|\tilde{g}(X) - g^*(X, W)\|^2_2}_{\text{Imputation Gap}} + \underbrace{\|g^*-Y\|^2_2}_{\text{Regression Noise}}\right\}, \label{eq:risk-decomp}
\end{align}
where $g^*(X,W) = \mathbb{E}[Y |X, W]$ is the optimal population surrogate. Then, by Lemma \ref{lemma:proximity} and \ref{lemma:imputation}, we obtain, with probability at least $1 - \left(\frac{1}{6} + \frac{1}{3}\right)e^{-x} = 1 - \frac{1}{2}e^{-x}$: 
\begin{align}
    \text{Var}(l_f) &\le 8B^2 \left( \epsilon^2 + C_2 \frac{\epsilon^{1-\alpha_{\mathcal{F}}/2}}{\sqrt{N^U}} + 2 L_g^2 \|W - \hat{W}\|_L^2 + C_3 \sqrt{\frac{\text{dim}(W)}{N^L}} + \frac{C_4}{\sqrt{N^L}} +\right. \notag \\
    &\quad \left. + 8B^2 \sqrt{\frac{x + \ln(6)}{2 N^U}} + (16B^2+8BL_g^2) \sqrt{\frac{x + \ln(6)}{2 N^L}} + \xi^2 \right) \coloneqq r(x), \label{eq:variance-final-modular}
\end{align}
where $\xi^2 = \|g^*-Y\|_2^2 \le B^2$ is the irreducible regression noise, $C_2 = \frac{16 B C \sqrt{C_{\mathcal{F}}}}{2-\alpha_{\mathcal{F}}}$,  $C_3 =  \frac{8L_g^2 \cdot C \sqrt{C_{\mathcal{Q}}'} (2B)^{1-\alpha_{\mathcal{Q}}/2}}{2-\alpha_{\mathcal{Q}}} $, and $C_4 =  \frac{16 C \sqrt{C_{\mathcal{G}}} \cdot (2B)^{1-\alpha_{\mathcal{G}}/2}}{2-\alpha_{\mathcal{G}}}$ are finite capacity constants.


\vspace{1em}
\textbf{Assembling the Theorem 1 Bound}. Finally, we assemble the concentration bounds. By applying the union bound over the Rademacher complexity of the localized function space \eqref{eq:bartlett-applied} (holding with probability at least $1 - \frac{1}{2}e^{-x}$) and the variance bound \eqref{eq:variance-final-modular} (holding with probability at least $1 - \frac{1}{2}e^{-x}$), with probability at least $1 - \left(\frac{1}{2} + \frac{1}{2}\right)e^{-x} = 1 - e^{-x}$. Every $f \in \mathcal{F}_\epsilon$ satisfies the uniform concentration bound:
\begin{align}
    \sup_{f \in \mathcal{F}_\epsilon} (\hat{R}_L(f) - R(f)) &\le 16B \cdot C_1 \frac{\epsilon^{1-\alpha_{\mathcal{F}}/2}}{\sqrt{N^L}} + \sqrt{\frac{2 r(x) x}{N^L}} + \frac{16 B^2}{3} \frac{x}{N^L}. \label{eq:assembled-concent-base}
\end{align}
To evaluate the square root term, we use the sub-additivity of the square root ($\sqrt{\sum a_i} \le \sum \sqrt{a_i}$). Specifically:
\begin{align}
    \sqrt{2r(x)} &\le 4B \Bigg( \epsilon + \sqrt{C_2} \frac{\epsilon^{1/2 - \alpha_{\mathcal{F}}/4}}{(N^U)^{1/4}} + \sqrt{2} L_g \|W - \hat{W}\|_L + \sqrt{C_3} \frac{\text{dim}(W)^{1/4}}{(N^L)^{1/4}} + \frac{\sqrt{C_4}}{(N^L)^{1/4}} + \xi  \notag \\
    &\quad + \sqrt{8B^2} \frac{(x + \ln 6)^{1/4}}{(N^U)^{1/4}} + \sqrt{16B^2+8BL_g^2}\frac{(x + \ln 6)^{1/4}}{(N^L)^{1/4}} \Bigg). \label{eq:r-sqrt-subadditivity}
\end{align}
Substituting the expanded square root bound \eqref{eq:r-sqrt-subadditivity} back into the concentration bound \eqref{eq:assembled-concent-base} requires evaluating the product of the tail term $(x+\ln 6)^{1/4}$ with the outer factor $\sqrt{x+\ln(6)} = (x+\ln 6)^{1/2}$, which yields $(x+\ln 6)^{3/4}$. To eliminate this fractional tail exponent and synchronize it with standard linear tail rates, we apply the Young's inequality (or AM-GM inequality) $u v \le \frac{3}{4} u^{4/3} + \frac{1}{4} v^4$ under the parameter choice $u = (x+\ln 6)^{3/4}$ and $v = (N^L)^{-3/4}$. This yields:
\begin{equation}
    \frac{(x+\ln 6)^{3/4}}{(N^L)^{3/4}} \le \frac{3}{4} \frac{x+\ln 6}{N^L} + \frac{1}{4 (N^L)^3}.
\end{equation}
Since $(N^L)^3 \ge N^L \ge 1$ for any sample size, this higher-order denominator term converges rapidly, allowing the fractional tail to be strictly upper bounded by a linear combination of $\frac{x+\ln 6}{N^L}$ and a constant. Absorbing these terms and grouping systematically by their common denominators yields the exact unnested uniform concentration bound:
\begin{align}
    \sup_{f \in \mathcal{F}_\epsilon} (\hat{R}_L(f) - R(f)) &\le 16B C_1 \left( \frac{\epsilon^{1-\alpha_{\mathcal{F}}/2}}{\sqrt{N^L}} \right) + 4B \sqrt{x + \ln(6)} \left( \frac{\epsilon + \sqrt{2} L_g \|W - \hat{W}\|_L + \xi}{\sqrt{N^L}} \right. \notag \\
    &\quad \left. + \sqrt{C_2} \frac{\epsilon^{1/2 - \alpha_{\mathcal{F}}/4}}{(N^U)^{1/4} \sqrt{N^L}} + \frac{\sqrt{C_3} \text{dim}(W)^{1/4} + \sqrt{C_4}}{(N^L)^{3/4}} \right) + C_5 \frac{x + \ln(6)}{N^L}, \label{eq:assembled-final-concentration}
\end{align}
where $C_5 > 0$ is the unified concentration tail constant. 

Combining \eqref{eq:assembled-final-concentration} directly with the local excess risk decomposition \eqref{eq:risk_limit_jasa}, and absorbing constants and the confidence parameter $x$ (and its square root $\sqrt{x}$) into the Big-O rate constants, the final excess risk of the ALCATRAs target predictive model converges under the following rates:
\begin{align}
    R(\hat{f}) - R(f^*) &\le \mathcal{O}\left( \frac{\epsilon^{1-\alpha_{\mathcal{F}}/2}}{\sqrt{N^L}} \right) + \mathcal{O}\left( \frac{\epsilon + L_g \|W - \hat{W}\|_L + \xi}{\sqrt{N^L}} \right) + \mathcal{O}\left( \frac{\epsilon^{1/2 - \alpha_{\mathcal{F}}/4}}{(N^U)^{1/4} \sqrt{N^L}} \right) \notag \\
    &\quad + \mathcal{O}\left( \frac{\text{dim}(W)^{1/4}}{(N^L)^{3/4}} \right) + \mathcal{O}\left( \frac{1}{N^L} \right) + \mathcal{O}(\epsilon^2).
\end{align}
This completes the proof of Theorem 1.
\end{proof}

\subsubsection{Lemma 2 (Lipschitz Property of Risks)}\label{sec:lipschitz_proof}
\begin{lemma}[Lipschitz Property of Risks]\label{lemma:lipschitz}
By Assumption~\ref{ass:boundedness}, the squared loss class $\mathcal{L} \circ \mathcal{F} := \left\{ (x, y) \mapsto (f(x) - y)^2 \mid f \in \mathcal{F} \right\}$ satisfies the Lipschitz property with constant $4B$. Consequently, by Talagrand's contraction principle:
\begin{equation}\mathbb{E}\mathscr{R}_n (\mathcal{L} \circ \mathcal{F}) \le 4B \cdot \mathbb{E}\mathscr{R}_n(\mathcal{F}).\end{equation}
\end{lemma}
\begin{proof}
Let $l(f(x), y) = (f(x) - y)^2$. For any two functions $f_1, f_2 \in \mathcal{F}$ and any $(x,y) \in \mathcal{X} \times \mathcal{Y}$, we evaluate the difference in their squared losses:
\begin{align*}
    |l(f_1(x), y) - l(f_2(x), y)| &= |(f_1(x) - y)^2 - (f_2(x) - y)^2| \\
    &= |(f_1(x) - y - (f_2(x) - y))(f_1(x) - y + f_2(x) - y)| \\
    &= |f_1(x) - f_2(x)| \cdot |f_1(x) + f_2(x) - 2y| \\
    &\le |f_1(x) - f_2(x)| \cdot (|f_1(x)| + |f_2(x)| + 2|y|).
\end{align*}
Under the boundedness assumption that $|f(x)| \le B$ and $|y| \le B$ for all $f \in \mathcal{F}$ (Assumption~\ref{ass:boundedness}), we have:
\begin{equation*}
    |f_1(x)| + |f_2(x)| + 2|y| \le B + B + 2B = 4B.
\end{equation*}
Thus, the loss function is $4B$-Lipschitz with respect to its first argument:
\begin{equation*}
    |l(f_1(x), y) - l(f_2(x), y)| \le 4B |f_1(x) - f_2(x)|.
\end{equation*}
Applying this contraction principle with $\phi_i(t) = (t - y_i)^2$, which has Lipschitz constant $L = 4B$ on $[-B, B]$, we directly obtain:
\begin{equation*}
    \mathbb{E}\mathscr{R}_n (\mathcal{L} \circ \mathcal{F}) \le 4B \cdot \mathbb{E}\mathscr{R}_n(\mathcal{F}),
\end{equation*}
which completes the proof.
\end{proof}

\subsubsection{Lemma 3 (Target-Surrogate Discrepancy Bound)}\label{sec:proximity_proof}
\begin{lemma}[Target-Surrogate Discrepancy Bound]\label{lemma:proximity}
Under Assumptions \ref{ass:boundedness}, \ref{ass:entropy}, \ref{ass:stability}, and \ref{ass:realizability}, with probability at least $1 - \frac{1}{6}e^{-x}$, the Target-Surrogate Discrepancy is bounded by:
\begin{equation}
    \|f - \tilde{g}\|_2^2 \le \epsilon^2 + C_2 \frac{\epsilon^{1-\alpha_{\mathcal{F}}/2}}{\sqrt{N^U}} + 8B^2 \sqrt{\frac{x + \ln(6)}{2 N^U}},
\end{equation}
where $C_2 = \frac{16 B C \sqrt{C_{\mathcal{F}}}}{2-\alpha_{\mathcal{F}}}$ is a finite capacity constant.
\end{lemma}
\begin{proof}
Applying McDiarmid's inequality to the empirical process $\sup_{f\in\mathcal F_\epsilon}(P-P_U)(f-\tilde g)^2$, with probability at least $1 - \frac{1}{6}e^{-x}$:
\begin{equation}
    \sup_{f \in \mathcal{F}_\epsilon} \left( \|f - \tilde{g}\|_2^2 - \|f - \tilde{g}\|_U^2 \right) \le 2 \mathbb{E}\mathscr{R}_{N^U}(\mathcal{H}) + 4B^2 \sqrt{\frac{x + \ln(6)}{2 N^U}}, \label{eq:f-g-concent-lemma3}
\end{equation}
Similar to Lemma \ref{lemma:lipschitz}, the expected Rademacher complexity satisfies $\mathbb{E}\mathscr{R}_{N^U}(\mathcal{H}) \le 4B \mathbb{E}\mathscr{R}_{N^U}(\mathcal{F}_\epsilon)$. Evaluating Dudley's entropy integral over the unlabeled sample size $N^U$ and empirical radius $\epsilon$ using the capacity bound in Assumption \ref{ass:entropy} yields:
\begin{equation}
    \mathbb{E}\mathscr{R}_{N^U}(\mathcal{F}_\epsilon) \le \frac{C}{\sqrt{N^U}} \int_0^\epsilon \sqrt{\log \mathcal{N}(\delta, \mathcal{F}_\epsilon, \|\cdot\|_{N^U})} \, d\delta \le \frac{2 C \sqrt{C_{\mathcal{F}}}}{2-\alpha_{\mathcal{F}}} \frac{\epsilon^{1-\alpha_{\mathcal{F}}/2}}{\sqrt{N^U}}.
\end{equation}
Substituting this Rademacher complexity back into \eqref{eq:f-g-concent-lemma3} and using $\|f - \tilde{g}\|_U^2 \le \epsilon^2$ uniformly completes the proof.
\end{proof}

\subsubsection{Lemma 4 (Completed Imputation Gap Bound)}\label{sec:imputation_proof}
\begin{lemma}[Completed Imputation Gap Bound]\label{lemma:imputation}
Under Assumptions \ref{ass:boundedness}, \ref{ass:entropy}, \ref{ass:lipschitz}, with probability at least $1 - \frac{1}{3}e^{-x}$, the Completed Imputation Gap satisfies:
\begin{equation}
    \|\tilde{g} - g^*\|_2^2 \le 2 L_g^2  \|W_i - \hat{W}_i\|_L^2 + C_3 \sqrt{\frac{\text{dim}(W)}{N^L}} + \frac{C_4}{\sqrt{N^L}} + (16B^2+8BL_g^2) \sqrt{\frac{x + \ln(6)}{2 N^L}}, \label{eq:lemma4-proven}
\end{equation}
where $C_3 =  \frac{8L_g^2 \cdot C \sqrt{C_{\mathcal{Q}}'} (2B)^{1-\alpha_{\mathcal{Q}}/2}}{2-\alpha_{\mathcal{Q}}} $, and $C_4 =  \frac{16 C \sqrt{C_{\mathcal{G}}} \cdot (2B)^{1-\alpha_{\mathcal{G}}/2}}{2-\alpha_{\mathcal{G}}}$ are constants.
\end{lemma}

\begin{proof}
Note that the completed surrogate function $\tilde{g}(X) = \hat{g}(X, \hat{W}(X))$ is constructed using the best imputed task features on the best surrogate estimator over the labeled dataset. Thus, to bound the completed imputation gap, we first decompose the population discrepancy into two terms:
\begin{equation}
    \|\tilde{g} - g^*\|_2^2 \le 2 \underbrace{\|\hat{g}(X,\hat{W}) - \hat{g}(X,W)\|_2^2}_{\text{Imputation error}} + 2 \underbrace{\|\hat{g}(X,W) - g^*(X,W)\|_2^2}_{\text{Surrogate generalization error}}. \label{eq:imputation-lipschitz-step}
\end{equation}
By Assumptions \ref{ass:lipschitz}, the online completed surrogate representation discrepancy satisfies: 
\begin{equation}
\|\hat{g}(X, \hat{W}) - \hat{g}(X, W))\|^2_2 \le L_g^2 \|W - \hat{W}\|_2^2 
\end{equation}
Then, applying the empirical process on the VAE reconstruction loss class $\mathcal{L}_{\text{VAE}} \coloneqq \{ (X, W) \mapsto \|W - \hat{W}_\theta(X)\|_2\}$ over the labeled sample set, with probability at least $1 - \frac{1}{6}e^{-x}$, we obtain:
\begin{equation}
    \|W - \hat{W}\|_2^2 \le \|W - \hat{W}\|_L^2 + 2 \mathbb{E}\mathscr{R}_{N^L}(\mathcal{L}_{\text{VAE}}) + 4B \sqrt{\frac{x + \ln(6)}{2 N^L}}, \label{eq:vae-uniform-concent-base}
\end{equation}
Bounding the expected Rademacher complexity using Dudley’s chaining integral over the variational class $\mathcal{Q}$ of $dim(W)$ output space yields:
\begin{align}
    \mathbb{E}\mathscr{R}_{N^L}(\mathcal{L}_{\text{VAE}}) &\le \frac{C}{\sqrt{N^L}} \int_0^{2B} \sqrt{\log \mathcal{N}(\delta, \mathcal{Q}, \|\cdot\|_{N^L})} \, d\delta \notag \\
    &\le \frac{C}{\sqrt{N^L}} \int_0^{2B} \sqrt{\text{dim}(W) \cdot C_{\mathcal{Q}}' \cdot  \delta^{-\alpha_{\mathcal{Q}}}} \, d\delta \\
    &= \left( \frac{2 C \sqrt{C_{\mathcal{Q}}'}}{2-\alpha_{\mathcal{Q}}} (2B)^{1-\alpha_{\mathcal{Q}}/2} \right) \sqrt{\frac{\text{dim}(W)}{N^L}} \label{eq:vae-rademacher-dudley}
\end{align}
Plugging in Equation \eqref{eq:vae-uniform-concent-base}, we have, with probability at least $1 - \frac{1}{6}e^{-x}$, the Imputation error is upper bounded by
\begin{equation}
    2 \cdot \|\Tilde{g} - \hat{g}\|^2_2 \le 2L_g^2 \|W - \hat{W}\|_L^2 + C_3\sqrt{\frac{\text{dim}(W)}{N^L}} + 8BL_g^2 \sqrt{\frac{x + \ln(6)}{2 N^L}}
    \label{eq:vae-l-concent}
\end{equation}
where $C_3 =  \frac{8L_g^2 \cdot C \sqrt{C_{\mathcal{Q}}'} \cdot (2B)^{1-\alpha_{\mathcal{Q}}/2}}{2-\alpha_{\mathcal{Q}}}$.

To evaluate the surrogate regression generalization, we analyze the global supremum of the empirical process over $\mathcal{G}$. Let $\hat{g} = \argmin_{g \in \mathcal{G}} \hat{R}_L(g)$ be the Empirical Risk Minimizer. By definition, $\hat{R}_L(\hat{g}) \le \hat{R}_L(g^*)$ holds, which yields the standard risk decomposition:
\begin{equation*}
    R(\hat{g}) - R(g^*) \le \left( R(\hat{g}) - \hat{R}_L(\hat{g}) \right) + \left( \hat{R}_L(g^*) - R(g^*) \right) \le 2 \sup_{g \in \mathcal{G}} |(\mathbb{P}_{N^L} - \mathbb{P})(g - g^*)|.
\end{equation*}
Under the squared loss, the population excess risk matches the $L_2(P)$ distance: $\|\hat{g}(X,W) - g^*(X,W)\|_2^2 = R(\hat{g}) - R(g^*)$. Applying the uniform concentration bound of Rademacher complexity to the global supremum over $\mathcal{G}$, with probability at least $1 - \frac{1}{6}e^{-x}$:
\begin{equation}
    \|\hat{g}(X,W) - g^*(X,W)\|_2^2 \le 4 \mathbb{E}\mathscr{R}_{N^L}(\mathcal{G}) + 8B^2 \sqrt{\frac{x + \ln(6)}{2 N^L}}. \label{eq:surrogate-global-sup-base}
\end{equation}
Evaluating Dudley's global entropy integral over the surrogate class $\mathcal{G}$ under Assumption \ref{ass:entropy}:
\begin{equation}
    \mathbb{E}\mathscr{R}_{N^L}(\mathcal{G}) \le \frac{C}{\sqrt{N^L}} \int_0^{2B} \sqrt{\log \mathcal{N}(\delta, \mathcal{G}, \|\cdot\|_{N^L})} \, d\delta \le \frac{2 C \sqrt{C_{\mathcal{G}}}}{2-\alpha_{\mathcal{G}}} \frac{(2B)^{1-\alpha_{\mathcal{G}}/2}}{\sqrt{N^L}}. \label{eq:surrogate-global-dudley}
\end{equation}
Substituting the global Dudley evaluation \eqref{eq:surrogate-global-dudley} back into \eqref{eq:surrogate-global-sup-base} yields:
\begin{equation}
    2 \cdot \|\hat{g}(X,W) - g^*(X,W)\|_2^2 \le \frac{C_4}{\sqrt{N^L}} + 16B^2 \sqrt{\frac{x + \ln(6)}{2 N^L}}, \label{eq:generalization-l}
\end{equation}
where $C_4 =  \frac{16 C \sqrt{C_{\mathcal{G}}} \cdot (2B)^{1-\alpha_{\mathcal{G}}/2}}{2-\alpha_{\mathcal{G}}} $.
Finally, by applying the union bound over \eqref{eq:vae-l-concent} and \eqref{eq:generalization-l}, we obtain the lemma results with probability at least $1 - 2 \cdot \frac{1}{6}e^{-x} = 1 - \frac{1}{3}e^{-x}$. 
\end{proof}

\subsection{Theoretical Properties of Streak Learning}

\subsubsection{Theorem2: Fisher Consistency of Streak Learning}
To analyze the theoretical properties of the proposed Streak Learning method, we first formalize the causal framework. Let $Y(\bm{a})$ be the potential labeling outcome if a sequence of tasks $\bm{a}$ is completed, and let $W_{\bm{a}}$ be the potential task outcomes. We assume the following standard sequential causal assumptions:

We recall the population Streak Value Function for a target policy $\bm{\pi} = (\pi_1, \dots, \pi_Q)$ with truncated optimal importance weights and post-streak optimal decisions indicator sum:
\begin{equation}
    \text{VF}^{\text{Streak}}(\bm{\pi}) = \mathbb{E} \left\{ \sum_{\delta=1}^{\bar{T}+1} \phi(\delta) \cdot \frac{R \cdot \prod_{t=1}^{\delta-1} \mathbb{I}(\pi_t(H_t)=A_t) \cdot \left( \sum_{t=\delta}^{\bar{T}} \mathbb{I}(\pi_t(H_t)=A_t) \right)}{\prod_{t=1}^T \pi^0_t(A_t \mid H_t)}\right\}.
\end{equation}

\begin{theorem}[Fisher Consistency of Streak Learning]\label{thm:fisher}
By Assumption~\ref{A-SUTVA}-\ref{A-pos} and the weighting function satisfies $\phi(\delta) > 0$ for all $\delta \in \{1, \ldots, Q+1\}$, the optimal policy $\bm{\pi}^*$ is the unique maximizer of the Streak Value Function:
\begin{equation}
    \bm{\pi}^* = \argmax_{\bm{\pi}} \text{VF}^{\text{Streak}}(\bm{\pi}).
\end{equation}
\end{theorem}
\begin{proof}[Proof of Theorem~\ref{thm:fisher}]
By the sequential ignorability and positivity conditions in Assumption~\ref{A-SUTVA}-\ref{A-pos}, we apply importance weighting up to stage $\delta-1$. which evaluates to the following hybrid expectation:
\begin{equation}
    \mathbb{E} \left\{ \frac{R \cdot \prod_{t=1}^{\delta-1} \mathbb{I}(\pi_t(H_t)=A_t) \cdot \left( \sum_{t=\delta}^{\bar{T}} \mathbb{I}(\pi_t(H_t)=A_t) \right)}{\prod_{t=1}^T \pi^0_t(A_t \mid H_t)}\right\} = \mathbb{E}^{\bm{\pi}_{1:\delta-1}, \bm{\pi}^0_{\delta:\bar{T}}} \left[ R \cdot \left( \sum_{t=\delta}^{\bar{T}} \mathbb{I}(\pi_t(H_t)=A_t) \right) \right].
\end{equation}
Thus, the Streak Value Function can be written as a weighted sum of expected hybrid rewards and optimal decisions under hybrid policies:
\begin{equation}
    \text{VF}^{\text{Streak}}(\bm{\pi}) = \sum_{\delta=1}^{\bar{T}+1} \phi(\delta) \mathbb{E}^{\bm{\pi}_{1:\delta-1}, \bm{\pi}^0_{\delta:\bar{T}}} \left[ R \cdot \left( \sum_{t=\delta}^{\bar{T}} \mathbb{I}(\pi_t(H_t)=A_t) \right) \right]. \label{eq:streak-hybrid-measure}
\end{equation}

By standard backward induction in sequential decision-making and dynamic programming optimality, at any intermediate stage $\delta \in \{1, \dots, Q+1\}$, the optimal sequence $\bm{\pi}^*$ uniquely and simultaneously maximizes the expected task reward $R$ and the number of optimal decisions made after the streak broke $\sum_{t=\delta}^{\bar{T}} \mathbb{I}(\pi_t(H_t)=A_t)$ under expectations:
\begin{equation}
    \mathbb{E}^{\bm{\pi}^*_{1:\delta-1}, \bm{\pi}^0_{\delta:\bar{T}}} \left[ R \cdot \left( \sum_{t=\delta}^{\bar{T}} \mathbb{I}(\pi^*_t(H_t)=A_t) \right) \right] \ge \mathbb{E}^{\bm{\pi}_{1:\delta-1}, \bm{\pi}^0_{\delta:\bar{T}}} \left[ R \cdot \left( \sum_{t=\delta}^{\bar{T}} \mathbb{I}(\pi_t(H_t)=A_t) \right) \right] \quad \forall \bm{\pi}. \label{eq:subgame-optimality}
\end{equation}
Since the weighting function satisfies $\phi(\delta) > 0$ strictly for all stages, we apply the inequality \eqref{eq:subgame-optimality} to each term in the summation \eqref{eq:streak-hybrid-measure}:
\begin{align*}
    \text{VF}^{\text{Streak}}(\bm{\pi}) &= \sum_{\delta=1}^{\bar{T}+1} \phi(\delta) \mathbb{E}^{\bm{\pi}_{1:\delta-1}, \bm{\pi}^0_{\delta:\bar{T}}} \left[ R \cdot \left( \sum_{t=\delta}^{\bar{T}} \mathbb{I}(\pi_t(H_t)=A_t) \right) \right] \\
    &\le \sum_{\delta=1}^{\bar{T}+1} \phi(\delta) \mathbb{E}^{\bm{\pi}^*_{1:\delta-1}, \bm{\pi}^0_{\delta:\bar{T}}} \left[ R \cdot \left( \sum_{t=\delta}^{\bar{T}} \mathbb{I}(\pi^*_t(H_t)=A_t) \right) \right] \\
    &= \text{VF}^{\text{Streak}}(\bm{\pi}^*).
\end{align*}
Since the inequality holds for any target policy $\bm{\pi}$ that deviates from the optimal policy $\bm{\pi}^*$ at any stage, we proved that $\bm{\pi}^*$ is the unique global maximizer of $\text{VF}^{\text{Streak}}(\bm{\pi})$.
\end{proof}


\subsubsection{Theorem3: Variance Boundedness and Sample Efficiency}
\begin{theorem}[Variance Dominance of Streak Learning]\label{thm:variance-reduction}
Under Assumptions \ref{ass:boundedness} and \ref{A-SUTVA}-\ref{A-pos} and denote $\bar{\delta} \coloneqq \mathbb{E}[\delta_i] \in [1, Q+1]$ be the expected number of streaks, then for any weighting function $\phi(t) > 0$, the variance of the Streak Learning estimator is strictly less than the standard IPW estimator:
\begin{equation}
    \text{Var}\left( \widehat{\text{VF}}^{\text{Streak}}(\bm{\pi}) \right) \le \mathcal{O}\left( \frac{B_R^2 Q^2 \phi(\bar{\delta})}{N^L \epsilon_0^{\bar{\delta}}} \right) \ll \text{Var}\left( \widehat{\text{VF}}^{\text{IPW}}(\bm{\pi}) \right) \asymp \mathcal{O}\left( \frac{B_R^2}{N^L \epsilon_0^Q} \right).
\end{equation}
\end{theorem}

\begin{proof}[Proof of Theorem~\ref{thm:variance-reduction}]
Because the Streak Learning and standard IPW estimators are constructed as empirical averages of $N^L$ independent, identically distributed $I_i \coloneqq \text{VF}_i$ evaluated at each off-policy trajectory observation, the sample variance of the value function equals:
\begin{equation}
    \text{Var}(\widehat{\text{VF}}) = \text{Var}\left( \frac{1}{N^L} \sum_{i=1}^{N^L} I_i \right) = \frac{\text{Var}(I)}{N^L}. \label{eq:clt-division-nl}
\end{equation}

We now calculate the exact variance of standard IPW. The standard IPW estimator is defined as:
\begin{equation}
    \widehat{\text{VF}}^{\text{IPW}}(\bm{\pi}) = \frac{1}{N^L} \sum_{i=1}^{N^L} I_{i,\text{IPW}} = \frac{1}{N^L} \sum_{i=1}^{N^L} \frac{R_i \prod_{t=1}^{Q} \mathbb{I}(\pi_t = A_{i,t})}{\prod_{t=1}^{Q} \pi^0_t(A_{i,t} \mid H_{i,t})}.
\end{equation}
The population second moment of a single standard IPW trajectory is:
\begin{align*}
    \mathbb{E}\left[ \left( I_{\text{IPW}} \right)^2 \right] &= \mathbb{E} \left[ \frac{R^2 \prod_{t=1}^{Q} \mathbb{I}(\pi_t = A_t)}{\prod_{t=1}^{Q} \pi^0_t(A_t \mid H_t)^2} \right] \\
    &= \mathbb{E}^{\bm{\pi}} \left[ \frac{R^2}{\prod_{t=1}^{Q} \pi^0_t(A_t \mid H_t)} \right].
\end{align*}
Using $\text{Var}(I_{\text{IPW}}) = \mathbb{E}[(I_{\text{IPW}})^2] - \mathbb{E}[I_{\text{IPW}}]^2$, and assuming the positivity condition $\pi^0_t(a_t \mid H_t) \ge \epsilon_0$ where the denominator is bounded below by $\epsilon_0^Q$, the empirical variance of standard IPW is calculated in a single intuitive Big-O chain as:
\begin{equation}
    \text{Var}\left( \widehat{\text{VF}}^{\text{IPW}}(\bm{\pi}) \right) = \text{Var}\left( \frac{1}{N^L} \sum_{i=1}^{N^L} I_{i,\text{IPW}} \right) = \frac{\text{Var}(I_{\text{IPW}})}{N^L} \asymp \mathcal{O}\left( \frac{B_R^2}{N^L \epsilon_0^Q} \right).
\end{equation}
This proves the exponential curse of horizon variance inflation.

Next, we analyze the Streak Learning estimator under the hybrid sequential decision formulation. The second moment of a single Streak Learning trajectory is:
\begin{equation}
    \mathbb{E}\left[ \left( I_{\text{Streak}} \right)^2 \right] = \mathbb{E} \left[ R^2 \left( \sum_{\delta=1}^{\bar{T}+1} \phi(\delta) \cdot \frac{\prod_{t=1}^{\delta-1} \mathbb{I}(\pi_t = A_t) \cdot \left( \sum_{t=\delta}^{\bar{T}} \mathbb{I}(\pi_t = A_t) \right)}{\prod_{t=1}^T \pi^0_t(A_t \mid H_t)} \right)^2 \right].
\end{equation}
Because $\bm{\pi}$ is a deterministic target policy, for any single observed trajectory $i$, let $\delta_i$ be the random stage where the target policy first deviates from the observed action sequence of trajectory $i$ (i.e., $\pi_t \neq A_t$ for $t = \delta_i$, and $\pi_t = A_t$ for all $t < \delta_i$).
Consequently:
\begin{equation}
    \prod_{t=1}^{\delta-1} \mathbb{I}(\pi_t = A_t) = 1 \quad \forall \delta \le \delta_i \;, \qquad \prod_{t=1}^{\delta-1} \mathbb{I}(\pi_t = A_t) = 0 \quad \forall \delta > \delta_i.
\end{equation}
This means that the Streak importance weighted sum inside the square truncates exactly at the deviation stage $\delta_i$:
\begin{equation}
    \sum_{\delta=1}^{\bar{T}+1} \phi(\delta) \cdot \frac{\prod_{t=1}^{\delta-1} \mathbb{I}(\pi_t = A_t) \cdot \left( \sum_{t=\delta}^{\bar{T}} \mathbb{I}(\pi_t = A_t) \right)}{\prod_{t=1}^{\delta-1} \pi^0_t(A_t \mid H_t)} = \sum_{t=1}^{\delta_i} \phi(t) \cdot \frac{\sum_{s=t}^{\bar{T}} \mathbb{I}(\pi_s = A_s)}{\prod_{r=1}^{t-1} \pi^0_r(A_r \mid H_r)}.
\end{equation}
which yields yields:
\begin{align}
    \mathbb{E}\left[ \left( I_{\text{Streak}} \right)^2 \right]  &= \mathbb{E}^{\bm{\pi}_{1:\delta-1}, \bm{\pi}^0_{\delta:T}} \left[ \frac{\phi(\delta)^2 R^2 \left( \sum_{t=\delta}^{\bar{T}} \mathbb{I}(\pi_t = A_t) \right)^2}{\prod_{t=1}^{\delta-1} \pi^0_t(A_t \mid H_t)} \right].
\end{align}
Applying the positivity condition $\pi^0_t \ge \epsilon_0$ yields the uniform lower bound for the first $\delta_i - 1$ stages: $\prod_{t=1}^{\delta_i-1} \pi^0_t(A_t \mid H_t) \ge \epsilon_0^{\delta_i-1} \ge \epsilon_0^{\delta_i}$. Substituting this and utilizing $\sum_{t=\delta}^{\bar{T}} \mathbb{I}(\pi_t = A_t) \le \bar{T} - \delta + 1 \le Q$:
\begin{equation}
    \mathbb{E}\left[ \left( I_{\text{Streak}} \right)^2 \right] \le B_R^2 Q^2 \cdot \mathbb{E}\left[ \frac{\phi(\delta_i)^2}{\prod_{t=1}^{\delta_i-1} \pi^0_t(A_t \mid H_t)} \right] \le B_R^2 Q^2 \cdot \mathbb{E}\left[ \frac{\phi(\delta_i)}{\epsilon_0^{\delta_i}} \right]. \label{eq:streak-second-moment-cancel}
\end{equation}
For any general decaying weighting function $0 < \phi(t) < 1$, and taking the expectation over the random deviation stage $\delta_i$ using Jensen's inequality under the expected deviation depth $\bar{\delta} \coloneqq \mathbb{E}[\delta_i]$, we obtain the closed-form second-moment bound:
\begin{equation}
    \mathbb{E}\left[ \left( I_{\text{Streak}} \right)^2 \right] \le B_R^2 Q^2 \cdot \frac{\phi(\bar{\delta})}{\epsilon_0^{\bar{\delta}}}. \label{eq:streak-second-moment-final}
\end{equation}
Plugging \eqref{eq:streak-second-moment-final} into \eqref{eq:clt-division-nl} yields the final closed-form variance dominance bound:
\begin{equation}
    \text{Var}\left( \widehat{\text{VF}}^{\text{Streak}}(\bm{\pi}) \right) \le \frac{B_R^2 Q^2 \phi(\bar{\delta})}{N^L \epsilon_0^{\bar{\delta}}}. \label{eq:streak-closed-form-variance}
\end{equation}
This completes the proof.
\end{proof}

\subsection{Causal Unbiasedness of Online-ELBO}

The main text (Section~4.3) presents a one-paragraph summary of this result and defers the full theorem statement and proof to this appendix. The motivation, stated briefly, is that under adaptive task acquisition, naive variational training of the surrogate task-outcome model is biased because the observed task outcomes are sampled by the evolving acquisition policy. The proposed Online-ELBO restores unbiasedness for the population variational objective by inverse-propensity weighting against the acquisition policy. Theorem~\ref{thm:elbo-unbiased} below formalizes this result, and the proof that follows shows that the policy probabilities cancel in expectation under positivity and sequential ignorability.
\begin{theorem}[Online-ELBO Unbiasedness]\label{thm:elbo-unbiased}
By Assumption~\ref{A-SUTVA}-\ref{A-pos}, the proposed Online-ELBO is causally unbiased for the true population Evidence Lower Bound of the overall task outcome distribution:
\begin{equation}
    \mathbb{E}_{X, A, W_A} [\text{ELBO}_{\text{online}}] = \text{ELBO}_{\text{population}}.
\end{equation}
\end{theorem}


\begin{proof}[Proof of Theorem~\ref{thm:elbo-unbiased}]
We evaluate the expectation of the reconstruction term in the Online-ELBO:
\begin{equation}
    E_{\text{recon}} = \mathbb{E}_{X, A, W_A} \left[ \frac{\mathbb{I}(A=q)}{\pi(A=q \mid X)} \log P(W_A \mid Z_q) \;\middle|\; Z_q \right]. \label{eq:recon-expectation}
\end{equation}
By the law of total expectation, conditioning on the observable features $X$:
\begin{equation*}
    E_{\text{recon}} = \mathbb{E}_{X} \left[ \mathbb{E}_{A, W_A \mid X} \left[ \frac{\mathbb{I}(A=q)}{\pi(A=q \mid X)} \log P(W_A \mid Z_q) \;\middle|\; X, Z_q \right] \right].
\end{equation*}
The inner conditional expectation over the observed actions $A$ and observed outcomes $W_A$ given $X$ is:
\begin{align*}
    &\mathbb{E}_{A, W_A \mid X} \left[ \frac{\mathbb{I}(A=q)}{\pi(A=q \mid X)} \log P(W_A \mid Z_q) \;\middle|\; X, Z_q \right] \\
    &\quad= \sum_{a} \pi(A=a \mid X) \mathbb{E}_{W_a \mid X, A=a} \left[ \frac{\mathbb{I}(a=q)}{\pi(A=q \mid X)} \log P(W_a \mid Z_q) \;\middle|\; X, Z_q \right] \\
    &\quad= \pi(A=q \mid X) \cdot \frac{1}{\pi(A=q \mid X)} \cdot \mathbb{E}_{W_q \mid X, A=q} \left[ \log P(W_q \mid Z_q) \;\middle|\; X, Z_q \right],
\end{align*}
where we used the fact that the indicator function $\mathbb{I}(a=q)$ is non-zero only when $a=q$. The selection probabilities cancel out perfectly:
\begin{equation*}
    = \mathbb{E}_{W_q \mid X, A=q} \left[ \log P(W_q \mid Z_q) \;\middle|\; X, Z_q \right].
\end{equation*}
Under the sequential ignorability assumption ($W_q \perp \!\! \perp A \mid X$), the potential outcome distribution of the task conditional on $X$ is independent of the policy's decision to assign that task. Thus, the conditional expectation matches the true population expectation:
\begin{equation*}
    \mathbb{E}_{W_q \mid X, A=q} \left[ \log P(W_q \mid Z_q) \;\middle|\; X, Z_q \right] = \mathbb{E}_{W_q \mid X} \left[ \log P(W_q \mid Z_q) \;\middle|\; X, Z_q \right].
\end{equation*}
Taking the outer expectation over $X$ and $Q_\phi(Z_q \mid X)$ in \eqref{eq:recon-expectation} yields the exact population reconstruction term:
\begin{equation*}
    \mathbb{E}_{X, W_q} \left[ \mathbb{E}_{Q_{\phi}(Z_q \mid X)} [\log P(W_q \mid Z_q)] \right].
\end{equation*}
Since the KL-divergence term in $\text{ELBO}_{\text{online}}$ only depends on $X$ (which is sampled from the true population distribution), the entire Online-ELBO is an unbiased estimator of the true population ELBO, i.e., $\mathbb{E}[\text{ELBO}_{\text{online}}] = \text{ELBO}_{\text{population}}$. This completes the proof.
\end{proof}

\section{Implementation and Algorithm} \label{chapter3-appendix: implementation}

This section presents the learning procedure of the proposed ALCATRAs framework under Streak Learning and online surrogate learning. The goal is to optimize four sets of parameters, $\theta, \psi, \phi$, and $\delta$, which parameterize the target predictive model, surrogate model, task variational model, and task-selection policy respectively. The complete iterative procedure is summarized in Algorithm~\ref{alg: AL-1}.

\begin{algorithm}[hbt!]
\caption{ALCATRAs Iteration}
\begin{algorithmic}[1]
\State \textbf{Input:} Unlabeled set $\mathbb{D}^U$, labeled set $\mathbb{D}^L$, budget $B$, uncertainty scoring function $U$, predictive model $f_\theta^0$, surrogate predictive model $g_\psi^0$, task-representation learner $Q_\phi^0$, task policy $\pi_\delta^0$, task-policy value function $V^\pi$, task reward set $R$.
\State \textbf{Output:} Updated labeled set $\mathbb{D}^L$, final predictive model $f$.
\State \textbf{Initialize:} iteration stage $t \gets 0$.
\While{budget $B$ is not exhausted}
    \State Select most uncertain unlabeled instance : $x_{\text{new}} = \arg\max_{x \in \mathbb{D}^{U}} U(f^{(t)}, x)$
    \State Determine an order of cost-effective tasks to be performed: $\bm{a}_{\text{new}} \gets \pi^{(t)}(x_{\text{new}})$
    \State Observe selected task outcomes and query the label for $x_{\text{new}}$, update data:
    $$
    \mathbb{D}^L \gets \mathbb{D}^L \cup \{x_{\text{new}}, (W_{\text{new}, \bm{a}_{\text{new}}}), y_{\text{new}}\}, \qquad \mathbb{D}^{U} \gets \mathbb{D}^{U} \setminus \{x_{\text{new}}\}
    $$
    \State (Su\parbox[t]{0.9\linewidth}{rrogate learning): update task-learner $Q^{(t+1)}_\phi$ by maximizing Equation \eqref{eq: online-ELBO}, impute missing task features and update the surrogate model $g^{(t+1)}  \gets \argmin_{g \in \mathcal{G}} \mathbb{E}_{|\mathbb{D}^L|}\{\mathcal{L}(g(X^L, \hat{W}^L), y^L)\}$, obtain surrogate labels and update model $f^{(t+1)}$ via Equation \eqref{eq: surr-eq}.}
    \State (Ta\parbox[t]{0.9\linewidth}{sk-policy learning): compute reward $R_{\text{new}}$ by Equation \eqref{eq:reward-concrete}, update the reward set $R \gets R \cup R_{\text{new}}$ and the task-selection policy $\pi^{(t+1)} \gets \argmax_\pi V^\pi(\mathbb{D}^L, R)$.}
    \State Update budget: $B \gets B - \text{cost}(x_{\text{new}}, W_{\text{new}, \bm{a}_{\text{new}}})$.
    \State Increase iteration stage: $t \gets t + 1$.
\EndWhile
\State \textbf{Return:} Updated labeled set $\mathbb{D}^L$ and final predictive model $f^{(t)}$.
\end{algorithmic}
\label{alg: AL-1}
\end{algorithm}

While the framework accommodates flexible model specifications, we parameterize both the target predictive model $f_\theta$ and the surrogate model $g_\psi$ as Multi-Layer Perceptrons (MLPs). The surrogate model $g_\psi$ augments the observable features $X$ with imputed task features $\hat{W}$, derived from the variational inference model when unobserved. For the task-selection policy $\pi_\delta$, we employ a fully-connected neural network architecture incorporating Long Short-Term Memory (LSTM) units to capture the temporal dependencies within the task acquisition history $H_t$. The sequence of observed features is mapped directly to a probability distribution over the available tasks, where the completed tasks are masked and the next task is selected via a greedy maximization of the policy output. The variational component is implemented as a Variational Autoencoder (VAE), where the encoder $Q_\phi(Z|X)$ and decoder $P(W|Z)$ are specified as MLPs and optimized via the reparameterization trick \citep{kingma2013auto}.

All the model parameters are optimized via Stochastic Gradient Descent (SGD) with momentum such as Adam \citep{kingma2014adam}. Due to the interdependent nature of the components, we adopt an iterative optimization scheme within each AL iteration. First, the variational parameters $\phi$ are updated to maximize the proposed Online-ELBO. Second, the surrogate parameters $\psi$ are updated using the imputed latent task features. Third, the target model parameters $\theta$ are refined using the augmented set of ground-truth and surrogate labels. Finally, based on the performance gains of the updated target model, we compute the rewards to update the task-selection policy for next AL iteration.

Hyperparameters, including the learning rate, batch size, and the number of hidden units for each network, are selected via $K$-fold cross-validation on the initial labeled dataset. To maximize data efficiency, the optimal configuration is chosen on the initial labeled data prior to the active acquisition phase, and remains fixed throughout the subsequent iterations.

\section{Closed-form simulation cost model} \label{chapter3-appendix: sim-cost}

This section provides the closed-form expressions for the simulation cost model summarized verbally in Section~6 of the main text. The full cost model decomposes as
\begin{equation}
    \text{costModel}(X, A_{1:\Tilde{T}}) = \underbrace{f^c(X) \vphantom{\textstyle \sum_{q \in A_{1:\Tilde{T}}}}}_{\text{Base costs}} \;+\; \underbrace{\textstyle \sum_{q \in A_{1:\Tilde{T}}} c_q}_{\text{Task costs}} \;+\; \text{EfficiencyCost}(A_{1:\Tilde{T}}; A^*_{1:\Tilde{T}^*}),
\end{equation}
where $f^c: \mathbb{R}^P \mapsto \mathbb{R}^+$ is a nonlinear function modeling the basic query cost and $c_q \sim \text{Unif}(0,1)$ is the per-task acquisition cost. The efficiency cost penalizes selecting lower-priority tasks before higher-priority ones (ranking penalty, scaled by $\alpha$) and measures strict deviation between the acquired and optimal task sequence (order penalty, scaled by $\beta$):
\begin{equation}
     \text{EfficiencyCost}(A_{1:\Tilde{T}}; A^*_{1:\Tilde{T}^*})
 = \underbrace{ \textstyle \vphantom{\left(1 - \frac{\sum_{t=1}^{|A^*|} \mathbb{I}(A_{\le t} = A^*_{\le t})}{|A^*|} \right)} \alpha \cdot
  \sum_{0 \le i < j \le \Tilde{T}} \frac{\mathbb{I}(\mathcal{O}^*(A_i) > \mathcal{O}^*(A_j))}{\binom{\Tilde{T}}{2}}}_{\text{Ranking costs}} + \underbrace{\textstyle \beta \cdot  \left(1 - \frac{\sum_{t=1}^{|A^*|} \mathbb{I}(A_{\le t} = A^*_{\le t})}{|A^*|} \right)}_{\text{Order costs}},
\end{equation}
where $\mathcal{O}^*$ maps the $A_i^{\text{th}}$ task to its ground-truth rank in the optimal selection sequence, and $\alpha, \beta \in [0,1]$ control the relative scale of the two penalties.

\section{FashionMNIST Synthetic Analysis} \label{chapter3-appendix: DA-fashion}

This section reports a parallel analysis to Section~7 of the main text on the FashionMNIST benchmark. The analysis is reported in the Supplementary Materials due to page constraints on the main text. The cost-weighted interpretation of the learned policy on UCI heart disease (Section~7 in the main text) is complemented here by a visually-interpretable analysis on FashionMNIST.

\begin{figure}[H]
    \centering
    \includegraphics[width=0.8\linewidth]{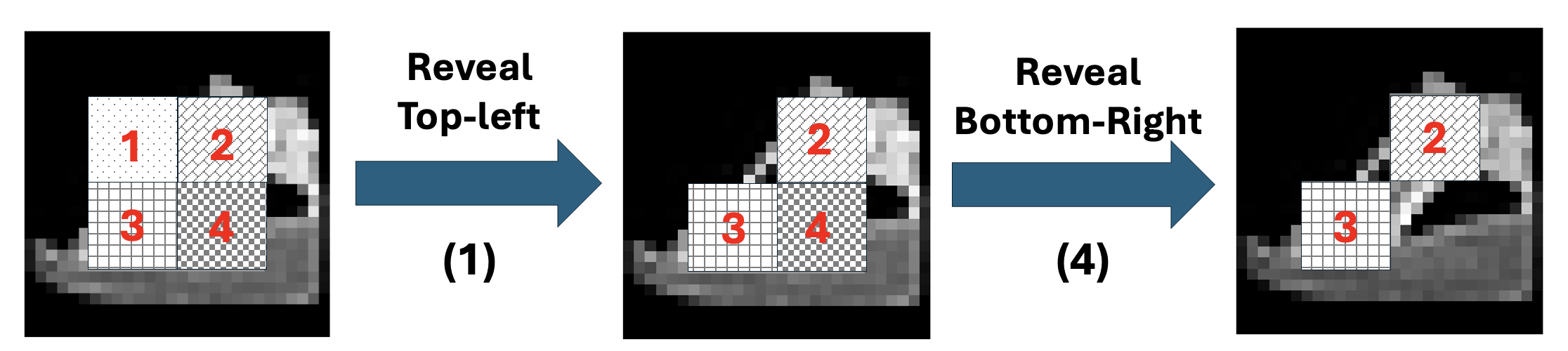}
    \caption{An example of synthetic tasks for FashionMNIST. This demonstrates a sequential selection of Task 1 followed by Task 4, where each corresponds to revealing a specific patch.}
    \label{fig:fashion-tasks}
\end{figure}

FashionMNIST is a benchmark dataset containing grayscale images of fashion items across 10 categories, such as shirts, sneakers, and handbags. The prediction task is to classify each image given its 28$\times$28 pixel representation. While the standard FashionMNIST does not involve additional labeling tasks, we generate synthetic tasks by masking the center of each image with four patches. Revealing each patch corresponds to a separate task and provides partial visual information of the image. Once all patches are revealed, the original image is fully recovered. An illustration of this patch-revealing process is shown in Figure~\ref{fig:fashion-tasks}.

\begin{figure}
    \centering
    \includegraphics[width=0.55\linewidth]{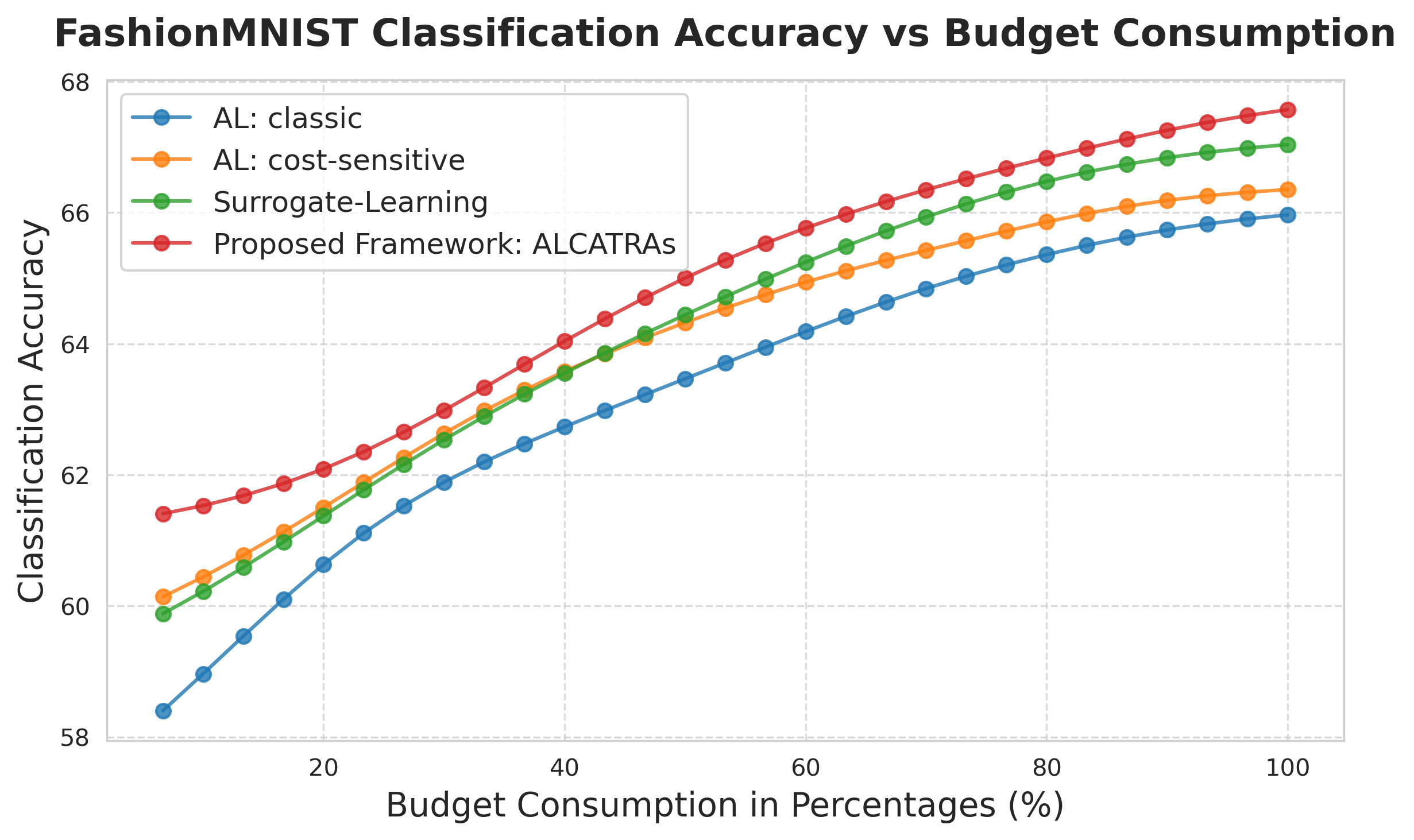}
    \caption{Trajectories of model predictive performance in classification accuracy versus budget consumption, averaged over 50 experiment repetitions. Each experiment includes a random selection of 150 labeled images, 2,000 unlabeled images, and 5,000 testing images. The maximum budget is 320 labor units, where querying the label of an unlabeled image costs 1 unit and revealing a single patch costs 0.25 units. A 100\% budget consumption means all budget has been utilized.}
    \label{fig:fashion-res}
\end{figure}

In this application, we train models on the fully-patched incomplete images and summarize the progression of model performance under different AL frameworks in Figure~\ref{fig:fashion-res}. The proposed ALCATRAs framework improves over baselines at every observed stage of budget consumption in this study. By leveraging partial visual information from patching tasks, ALCATRAs achieves a strong initial performance improvement from high-quality surrogate labels. In addition, while ALCATRAs improves relatively slowly during the early budget phase (0\%--20\%) as the task-selection policy is still learning, it quickly adapts once the policy improves and achieves the best performance onwards. These results demonstrate the essential role of task information in the learning procedure, which, however, is currently missing in both surrogate and cost-sensitive learning despite their improvements over the classic AL. We also remark that integrating task information directly into surrogate labels is more efficient than modeling separate belief states as in the feature acquisition framework, whose results are not comparable and omitted here due to the challenges of modeling belief states from high-dimensional images with limited samples.

\begin{figure}[hbt!]
    \centering
    \includegraphics[width=0.75\linewidth]{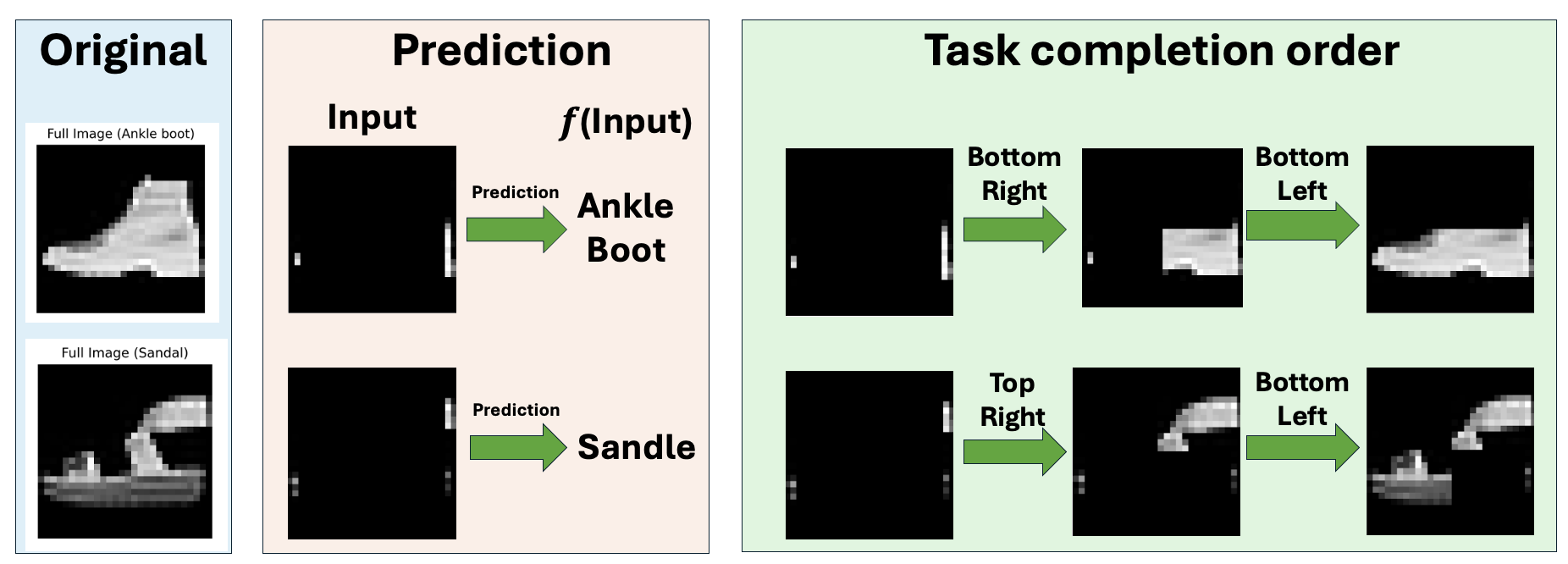}
    \caption{Anecdotal analysis of the learned FashionMNIST acquisition decision rule.}
    \label{fig:chapter3-appendix: annecdote}
\end{figure}

Next, to better interpret the obtained acquisition decision rule, among the 10 fashion item categories, we focus on sandals and boots, and randomly select one illustrative example from each category for anecdotal analysis. The original (fully revealed) images are displayed in Figure~\ref{fig:chapter3-appendix: annecdote}, and the trained classification model successfully predicts the correct class labels from the partially revealed images in both examples. When examining the task completion order, the learned acquisition rule shows distinct selection patterns across these two categories. For boots, it tends to first reveal the bottom-right patch followed by the bottom-left. In contrast, for sandals, the rule prioritizes the top-right patch, followed by the bottom-left patch. This order is intuitively interpretable: boots are most recognizable by their heels, located at the bottom of the images, while sandals commonly feature cut-out designs along the diagonal from toes to the upper side of the shoe. As an anecdotal observation, the acquisition rule appears to align with intuitive visual reasoning and to assist the classifier with similarly interpretable behavior.

\addtocontents{toc}{\protect\setcounter{tocdepth}{2}}

\newpage 
{
\small
\bibliography{reference}
}